# The emergence of Large Language Models (LLM) as a tool in literature reviews: an LLM automated systematic review

Dmitry Scherbakov, PhD[1,*], Nina Hubig, PhD [1,2,*], Vinita Jansari, PhD[2], Alexander Bakumenko, MSc[2] , Leslie A. Lenert, MD[1]

[1] Biomedical Informatics Center, Department of Public Health Sciences, Medical University of South Carolina (MUSC), Charleston, South Carolina, USA.

[2] Clemson University, School of Computing , Charleston, South Carolina, USA.

* Authors with equal contribution.

**Corresponding author:** Leslie Lenert (lenert@musc.edu)
**Address:** 22 WestEdge Street, Suite 200, Room WG213, Charleston, South Carolina, 29403, USA

**Manuscript word count: 2987** (excluding title page, abstract, references, tables, acknowledgements, funding, data availability, competing interests, and author contributions statements)
**Abstract word count: 301** (excluding keywords)

Highest academic degrees of authors (in order):
DS : PhD, Postdoctoral scholar
NH : PhD, Assistant professor
VJ : PhD, Postdoctoral scholar
AB : MSc, Doctoral student
LL : MD, MS, FACP, FACMI, Professor

## Short description

This study aims to summarize the usage of Large Language Models (LLMs) in the process of creating a scientific review. The idea behind this publication is to turn the tool “on itself”, conducting a systematic review of research projects using LLMs for systematic and other types of reviews by using a set of LLM tools.

**Abstract**

**Objective**: This study aims to summarize the usage of Large Language Models (LLMs) in the process of creating a scientific review. We look at the range of stages in a review that can be automated and assess the current state-of-the-art research projects in the field.

**Materials and Methods**: The search was conducted in June 2024 in PubMed, Scopus, Dimensions, and Google Scholar databases by human reviewers. Screening and extraction process took place in Covidence with the help of LLM add-on which uses OpenAI gpt-4o model. ChatGPT was used to clean extracted data and generate code for figures in this manuscript, ChatGPT and Scite.ai were used in drafting all components of the manuscript, except the methods and discussion sections.

**Results**: 3,788 articles were retrieved, and 172 studies were deemed eligible for the final review. ChatGPT and GPT-based LLM emerged as the most dominant architecture for review automation (n=126, 73.2%). A significant number of review automation projects were found, but only a limited number of papers (n=26, 15.1%) were actual reviews that used LLM during their creation. Most citations focused on automation of a particular stage of review, such as Searching for publications (n=60, 34.9%), and Data extraction (n=54, 31.4%). When comparing pooled performance of GPT-based and BERT-based models, the former were better in data extraction with mean precision 83.0% (SD=10.4), and recall 86.0% (SD=9.8), while being slightly less accurate in title and abstract screening stage ($M_{accuracy}$=77.3%, SD=13.0 vs $M_{accuracy}$=80.9% SD=11.8).

**Discussion/Conclusion**: Our LLM-assisted systematic review revealed a significant number of research projects related to review automation using LLMs. The results looked promising, and we anticipate that LLMs will change in the near future the way the scientific reviews are conducted, significantly reducing the time required to generate systematic reviews of the literature and expanding how systematic reviews are used to guide science.

## Introduction

The abundance of scientific information available can be overwhelming, posing a challenge for researchers to navigate relevant data. Consequently, scoping and systematic reviews that are helping scientists synthesize the evidence have seen a *significant increase* over the years. Toh & Lee noted an exponential rise in the number of scoping reviews, with 2,665 scoping reviews being published in 2020 alone, compared to less than 10 reviews annually before 2009 [1]. The same trend is observed in systematic reviews and meta-analyses, for example, in cardiology over 2,400 meta-analyses were published in 2019, quadruple the number from 2012 [2].

A completion of a review requires substantial resources; further, there is often unpredictable uncertainty in the amount of resources required [3]. The time to complete a single systematic review varies, but authors typically give estimates in months and even years [4]. Screening automation platforms, such as Covidence [5], facilitate systematic and scoping reviews by streamlining established guidelines, such as the Preferred Reporting Items for Systematic Reviews and Meta-Analyses (PRISMA) and PICO (Population, Intervention, Comparison, and Outcome) to ensure transparency and rigor in the review process [6]. The use of such platforms may reduce the time to complete reviews by providing tools that automate key tasks, such as removing duplicate references, generating flow-charts of the screening process, visual extraction designers, and workflows for several independent reviewers.

Although, for example, Covidence, includes features to reduce the time to complete screening, such as key term highlighting and embedded natural processing (NLP) algorithm [7] it primarily organizes the significant manual work that is still needed from human reviewers like screening and extraction.. Each of these steps normally requires two independent analysts, with a third optional human expert supervising the process and resolving the disagreements.

Even with two reviewers double-checking each other, as much as 3% of relevant citations are missed, and if only a single reviewer is used (for example, in rapid reviews), as many as 13% of relevant publications can be missed [8]. The relatively weak performance of humans in screening relevant articles has led some investigators to develop natural language processing tools [9-12] to automate screening. A recent statement by the National Institute for Health and Care Excellence (NICE) highlights a big potential of AI in the systematic review process automation [13].

Large Language Models (LLM) recently emerged as one of the most powerful NLP tools across different ranges of tasks. By conducting this review, we wanted to evaluate the natural extension of the use of LLMs to guide and direct the review process. Thus, this systematic review aims to (1) summarize the current state-of-the-art research projects using LLMs to automate the review process, (2) look at the range of review types and review stages that are being automated, (3) assess the quality of each research project, (4) assess the performance of LLMs used for automation.

As LLMs are used as a possible substitute for a human reviewer, the idea behind this publication is to turn the tool "on itself", conducting a systematic review of research projects using LLMs for systematic reviews.

**Methods**

The study's research plan was formulated by the author team and adjusted based on the guidance provided by the preferred reporting items for systematic review and meta-analysis protocols (PRISMA-P) 2015: elaboration and explanation [14] and the latest JBI checklist [15] for conducting systematic reviews. The review protocol was registered in the Open Science Framework (OSF) database [16].

We decided that to be included in the review, citations had to be centered around the usage of LLMs in automation of different phases of systematic review. Only English-language journal publications were considered, including, conference abstracts, and review publications that used LLMs in their creation.

Publications were excluded if they:

- Did not use some kind of LLM (e.g. ChatGPT, Mistral, GPT-3.5, BERT)
- Did not describe automation of any stage of the review process
- The paper was a review article itself that did not use LLM to conduct the review
- Full text of the article could not be retrieved or was not in English

The initial search was conducted by a human reviewer (DS) in June 2024 in PubMed, Scopus, Dimensions, and Google Scholar databases. Table 1 presents the search strategy for the databases.

**Table 1. Search strategy.**

(("large language models" OR "large language model" OR "LLM" OR "LLMs" OR "ChatGPT" OR "GPT-3" OR "GPT-4" OR "LLaMA" OR "Mistral" OR "Mixtral" OR "BARD" OR "BERT" OR "Claude" OR "PaLM" OR "Gemini" OR "Copilot") **AND** ("systematic review*" OR "scoping review*" OR "literature review*" OR "narrative review*" OR "umbrella review*" OR "rapid review*" OR "integrative review*" OR "evidence synthesis" OR "meta-analysis"))

*Source: Authors' own work*

All citations were then uploaded to Covidence. The screening and extraction process took place in Covidence with the help of LLM plugin for Covidence that our team developed. This plugin is used during screening and extraction phases. The process of using LLM for screening and extraction is shown on Figure 1.

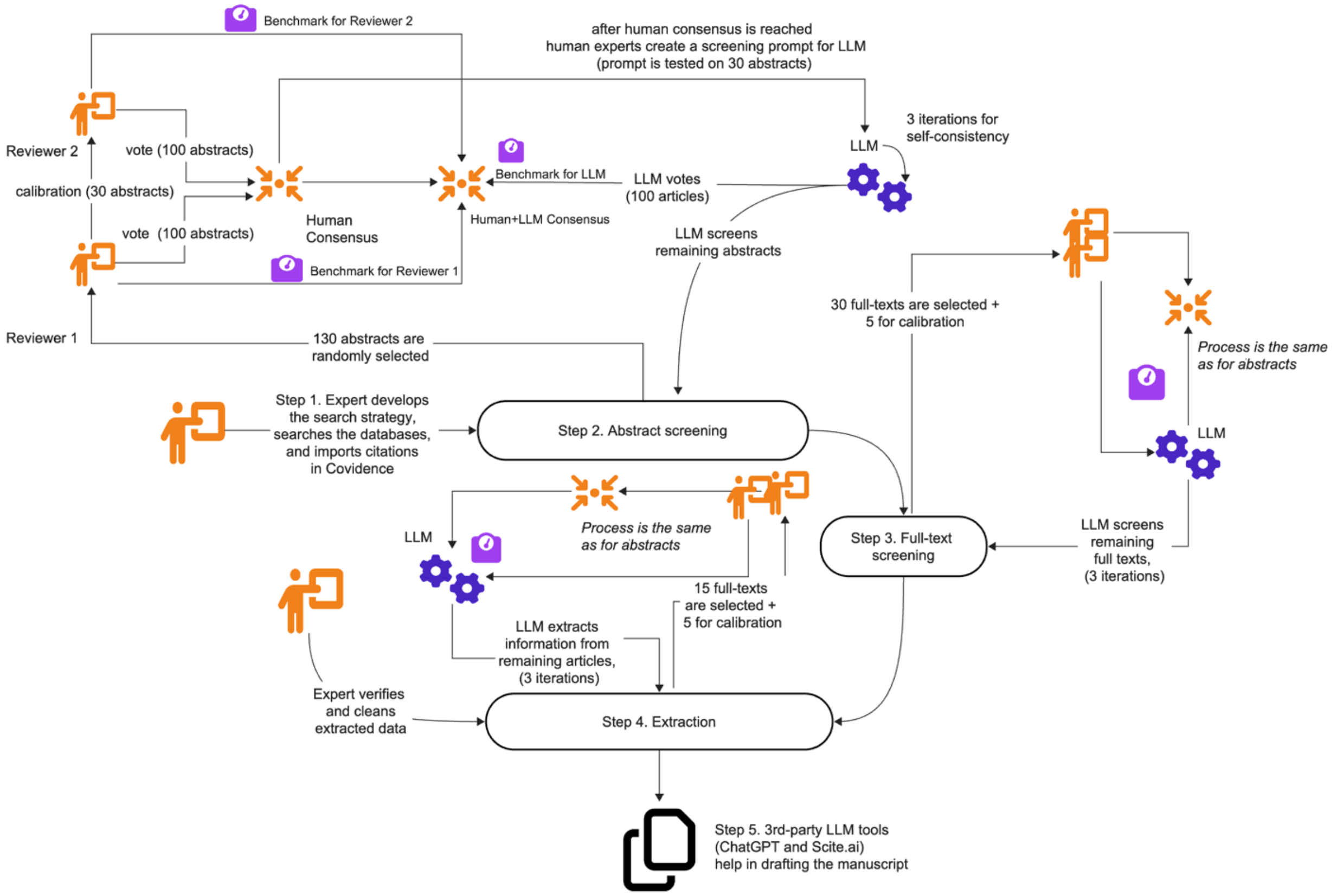

**Figure 1.** LLM workflow added into Covidence for screening and extraction.

*Source: Authors' own work.*

The developed add-on works by interacting with the Covidence platform programmatically via an intermediary software solution that was created in Python and R. The solution passes contents between Covidence and the LLM OpenAI gpt-4o model . Once the LLM generates the response, a script automates actions in Covidence, such as clicking Include/Exclude buttons or leaving notes.

The review process involved three stages that were automated by Covidence add-on: abstract screening, full-text screening, and extraction. In each stage, two human reviewers calibrated by screening a sample to refine inclusion criteria and extraction categories. They then created and tested prompts for the LLM. LLM inference was programmed to run inference 3 times to determine the final decision (e.g., "include" or "exclude") based on the majority vote. Three prompts per phase are detailed in Supplementary Appendix S1.

For abstract screening, LLM and human reviewers voted for consensus, and a human expert consensus was established. In full-text screening and extraction, a single human reviewer verified LLM results. Extraction precision was measured, and for categories with low precision (<80%), a manual reviewer validated LLM outputs. Benchmarks are provided in Supplementary Appendix S2.

The data charting form for extraction were designed by human experts (DS, VJ, AB, LL, and NH) and adopted into the LLM prompt to collect the following primary information:

- Author, year, title;
- Country and/or US state;
- What types of reviews were automated;
- Stage of review automated in the research project;
- LLM type used;
- Performance metrics reported by authors during each stage of the review. In particular, Accuracy, Precision, Recall, Specificity, and F1 were extracted, if other metrics were used instead, they were grouped under "Other metrics" category;
- Brief information on how were these performance metrics calculated;
- Brief information on reported timesaving;
- What was general opinion of the study team on the usage of LLMs in review automation (positive, negative, or mixed) with a citation to support this viewpoint.

Human reviewers (DS, VJ, AB) performed quality assessment of given studies using a set of selected categories from the reviewed studies and a points-based scale:

- Ratings of the universities where authors are affiliated (the data was linked to QS ranking 2024 [17]), maximum value across all co-author affiliations was used:
    - Ranked 1 to 100: 2 points
    - 100-1000: 1 point
    - >1000: 0 point
- Number of samples (full-texts or abstracts) that authors used to compute their performance metrics:
    - More than 200: 2 points
    - 50-200: 1 point
    - <50: 0 points
- Sources of the funding of the research project (public, private or mixed)
    - Public funding: 2 points
    - No funding: 1 point
    - Private funding: 0 points
- Impact factor of the journal [18]
    - More than 5: 2 points
    - 1 to 5: 1 point
    - Less than 1: 0 points
- Is the paper an actual review which used LLM?
    - A review: 2 points
    - Not a review (methods paper): 1 point
- Were performance metrics (benchmarks) reported?
    - 2 points for reporting performance metrics
    - 0 points for no metrics

If value in any above category could not be determined (e.g. no match for university or impact factor, or unknown value in category), then the NA value was assigned. Based on the mean of points across all the quality categories, studies were classified as low (<1 points), medium (1 to 1.5 points) or high quality (>=1.5 points).

An LLM tool by Google (NotebookLM, version from August 2024) along with a manual review (DS, VJ, AB) was used to cross check the extraction results for the fields where precision of extraction was low (<0.8) during the benchmark. Again, ChatGPT (4o model) was used to clean the extraction data: format the case, remove duplicates, rename similar entries to a common name. The data was then manually fed into the chat window by a human reviewer (DS). Scite.ai (version from August 2024) was used to draft parts of the introduction and discussion sections, while ChatGPT was used to draft the abstract and results section of this review by generating R code snippets to produce all figures (except Figure 1 which was generated by Covidence). ChatGPT was also used to draft the text of the results section, which was then corrected by our team where needed. Human experts edited and verified the final LLM-generated draft of the manuscript.

Additionally, we report the time saving and the computational costs in Supplementary Appendix S3. We used our own time measurements and reference data from experienced reviewers to calculate time-saving [19].

**Results**

Figure 2 outlines the PRISMA article selection process for this study. Initially, 3,788 studies were identified across several databases: PubMed (n = 2,174), Scopus (n = 1,207), Dimensions (n = 356), and Google Scholar (n = 48), along with 3 additional studies from citation searching. Following the removal of 447 duplicates (1 manually and 446 by Covidence), 3,341 studies remained for the screening phase.

During the title and abstract screening process, 3,041 studies were excluded, leaving 300 studies for retrieval and full-text eligibility assessment. Out of these 300 studies, 128 were excluded for various reasons, with the most common being “The paper does not describe the automation of any stage of the review process” (n = 88). A total of 172 studies were included in the final review.

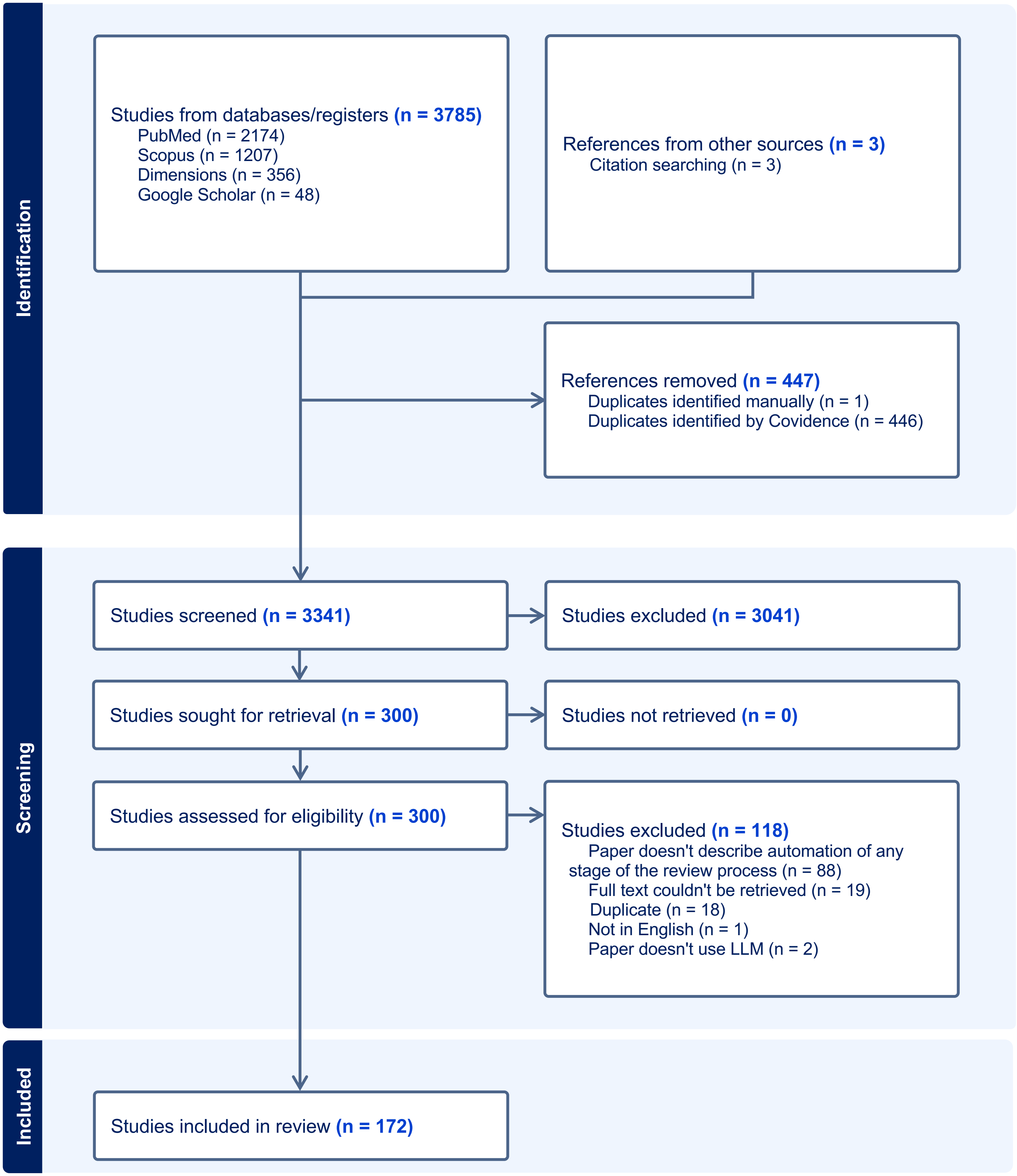

**Figure 2.** Flow diagram of the systematic review process. *Source: Authors' own work / Covidence.*

Figure 3 shows the geographic distribution of studies across 43 countries. Most citations are from the US (n=60, 34.9%), followed by Australia (n=14, 8.14%), the UK and China (n=13, 7.6%), and Germany (n=11, 6.4%). Other notable contributors include Canada (n=7, 4.1%) and India (n=6, 3.5%). Austria, Ireland, Italy, the Netherlands, and South Korea each contributed 4 studies (2.3%), while countries like New Zealand, France, Japan, and others provided 3 (1.7%). The rest contributed 1–2 studies.

In the US, 47 studies had state-level data. Tennessee, New York, and Massachusetts led with 5 citations each (10.6%), followed by California (n=4, 8.5%). North Carolina and Ohio contributed 3 studies (6.4%), while several other states provided 2 (4.3%) or 1 (2.1%) citation

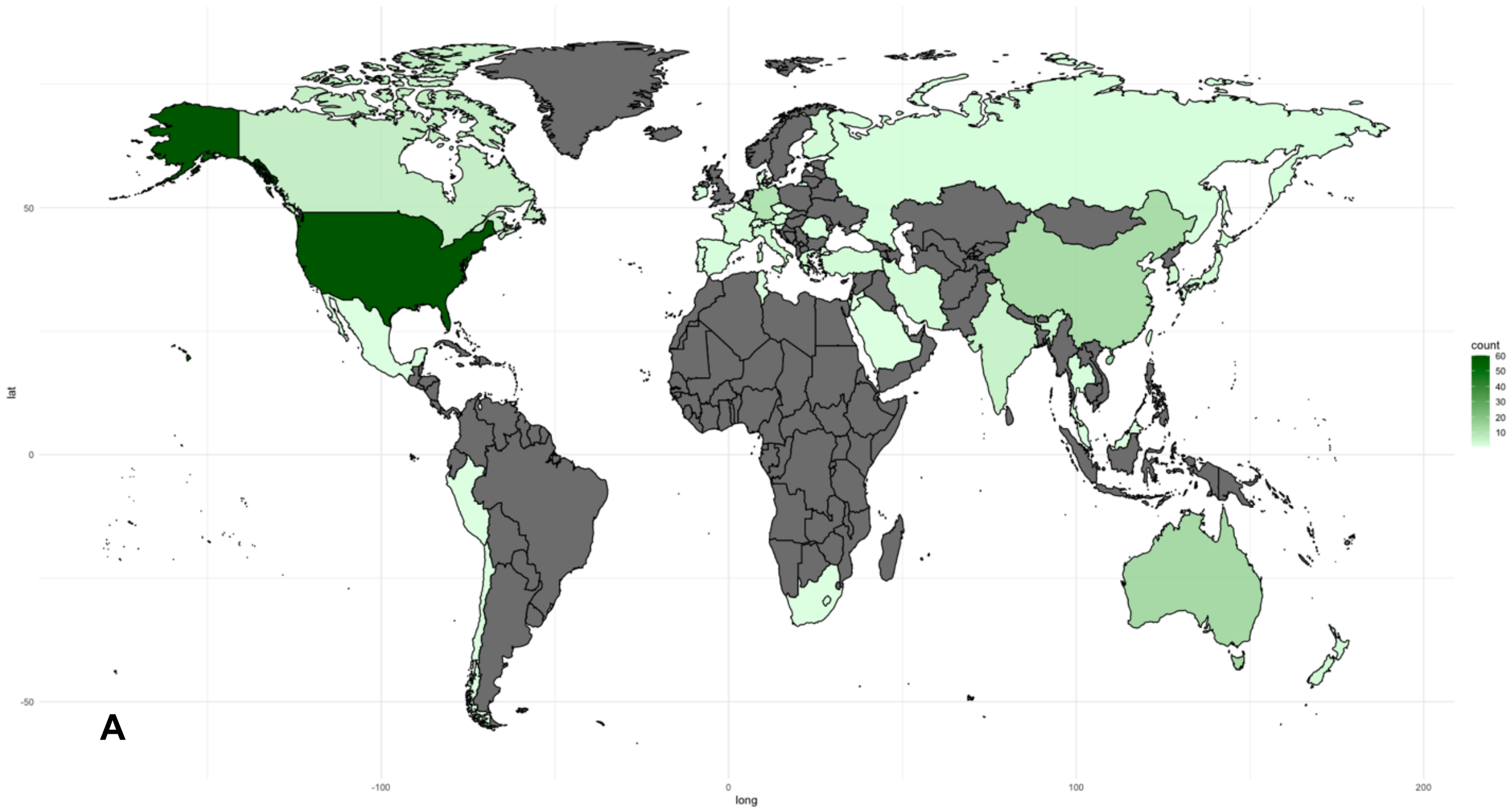

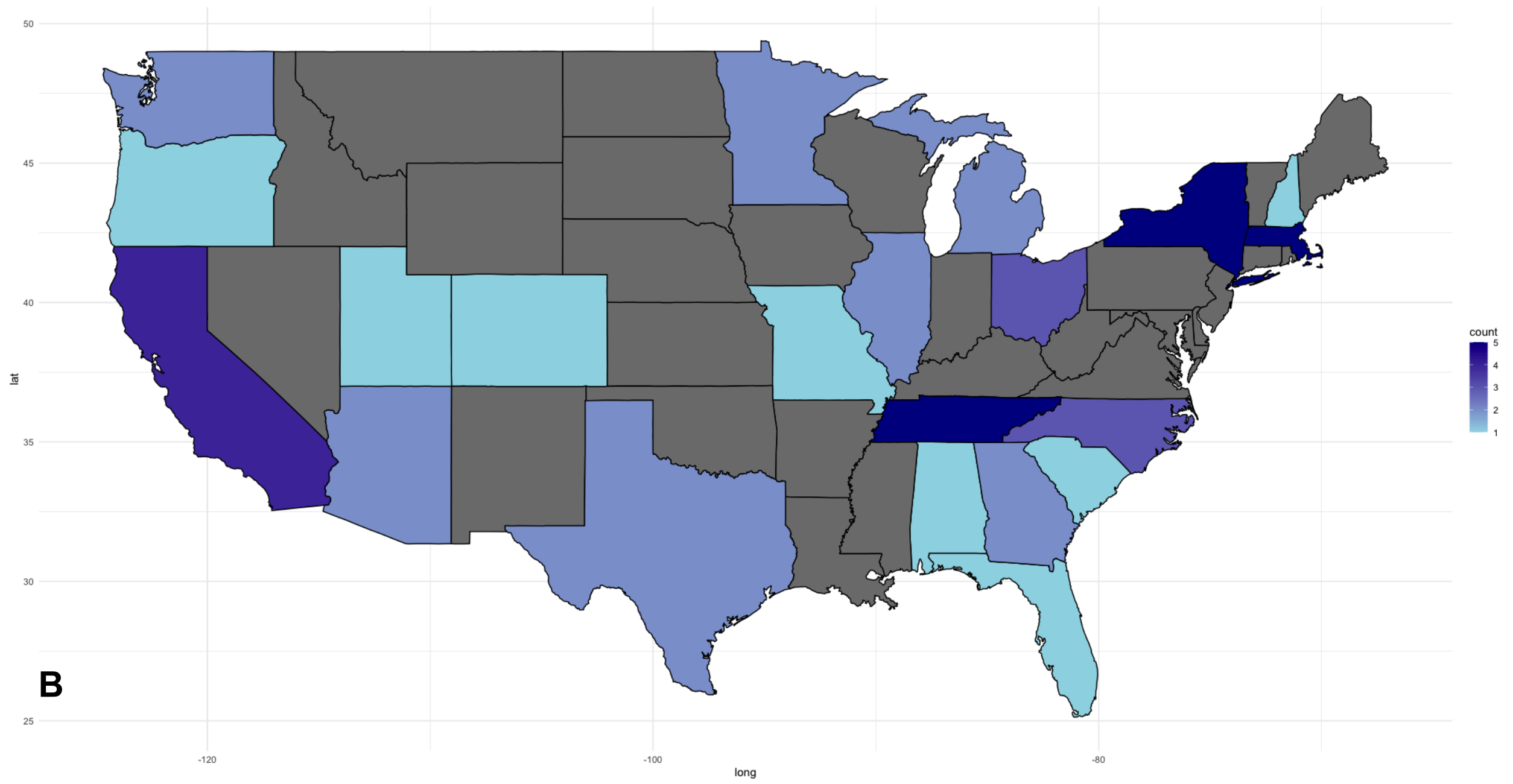

**Figure 3. A:** Publications by country of origin; **B:** Publications by state in the US. *Source: ChatGPT-generated code using extracted data from this manuscript (Supplemental Table S7)*

Figure 4A shows the types of reviews discussed in automation papers. The most frequently mentioned type is 'Systematic Review' (n=118, 68.6%), followed by 'Literature/Narrative Review' (n=37, 21.5%) and 'Meta-Analysis' (n=19, 11.0%). The remaining categories include 'Scoping Review' (n=8, 4.7%), 'Other/Non-specific' (n=14, 8.1%), and 'Rapid Review' (n=6, 3.5%). 'Umbrella Review' has a smaller representation with 2 mentions (1.2%).

Figure 4B illustrates the stages of review discussed in automation papers. The most frequently mentioned stage is 'Searching for publications' (n=60, 34.9%), followed by 'Data extraction' (n=54, 31.4%) and 'Evidence synthesis/summarization' (n=32, 18.6%). Other categories with notable mentions include 'Title and abstract screening' (n=43, 25.0%), 'Drafting a publication' (n=22, 12.8%), 'Full-text screening' (n=14, 8.1%), 'Quality and bias assessment' (n=12, 7.0%), 'Publication classification' (n=10, 5.8%), 'Other stages' (n=6, 3.5%), and 'Code and plots generation' (n=4, 2.3%).

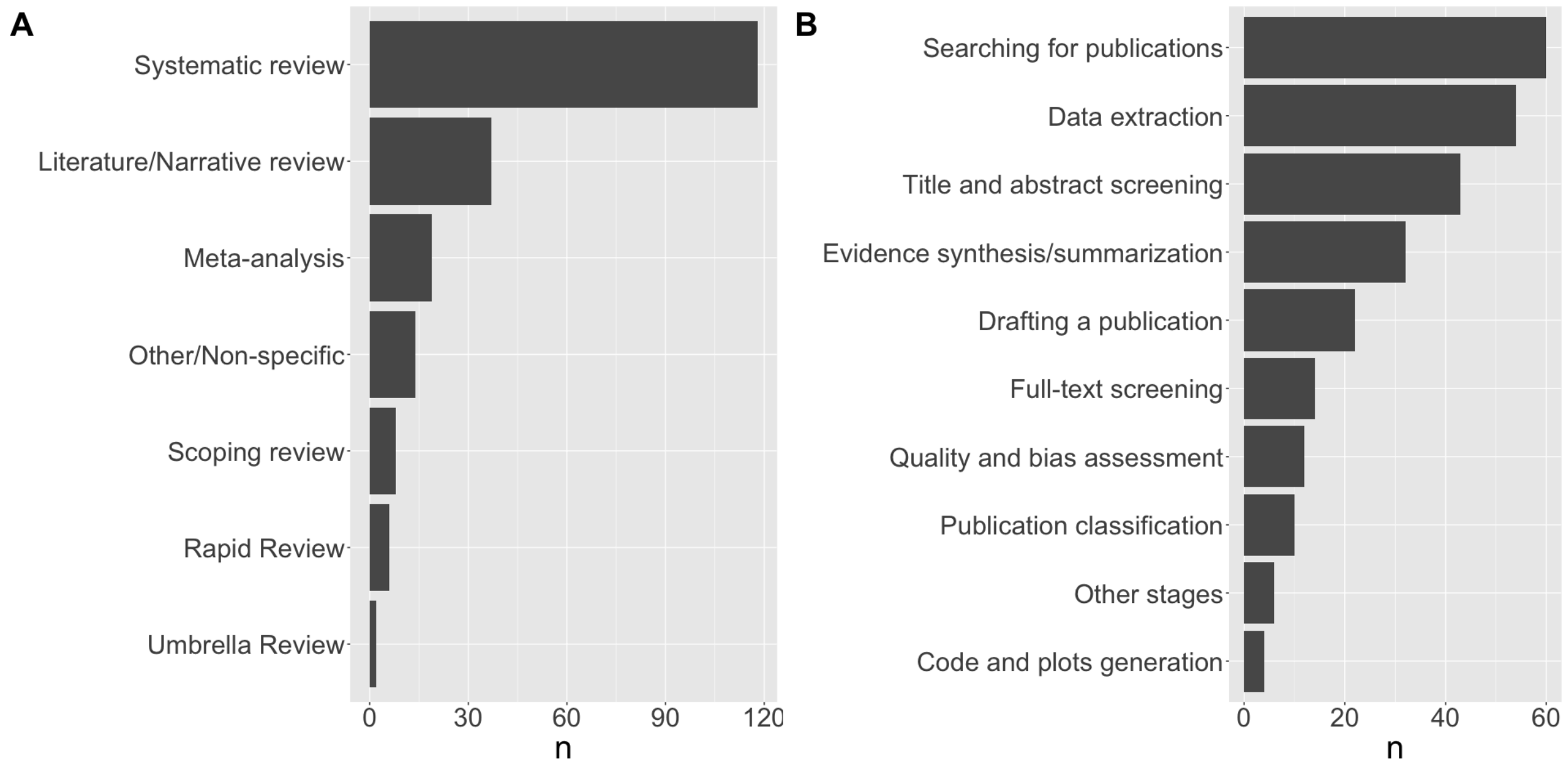

**Figure 4.** A. Types of review automated B. Which stages of review are automated in the paper. *Source: ChatGPT-generated code using extracted data from this manuscript (Supplemental Table S7)*

The most frequently mentioned AI model is GPT/ChatGPT, with 126 occurrences (73.3%), showing its widespread use (Supplemental Figure S5). BERT-based models are also notable with 32 mentions (18.6%). LLaMA/Alpaca models have 8 mentions (4.7%), followed by Google Bard/Gemini with 5 (2.9%), and Claude models with 7 (4.1%). Other models like BART (n=3, 1.7%) and Mistral (n=4, 2.3%) are less frequent. Several models, including Bing and XLNet, have 2 mentions each (1.2%), while many others are mentioned just once (0.6%).

Of the 172 citations, 79 (45.9%) reported common metrics like Accuracy, Precision/Recall, and F1, while 36 (20.9%) used less common metrics like G-score and Jaccard similarity. The remaining 57 publications (33.1%) relied on qualitative assessments.

Figure 5 shows performance metrics for GPT- and BERT-based models. GPT models had lower accuracy in title/abstract screening (M=77.34, SD=13.06) compared to BERT models (M=80.87, SD=11.81). However, GPT models performed better in data extraction, with precision (M=83.07, SD=10.43) and recall (M=85.99, SD=9.82), while BERT models had lower precision (M=61.06, SD=31.26) and similar recall (M=80.03, SD=10.09). In title/abstract screening, BERT models had higher precision (M=65.6, SD=17.65) but lower recall (M=72.93, SD=23.95) than GPT models (precision M=63.2, SD=24.34; recall M=80.42, SD=23.31).

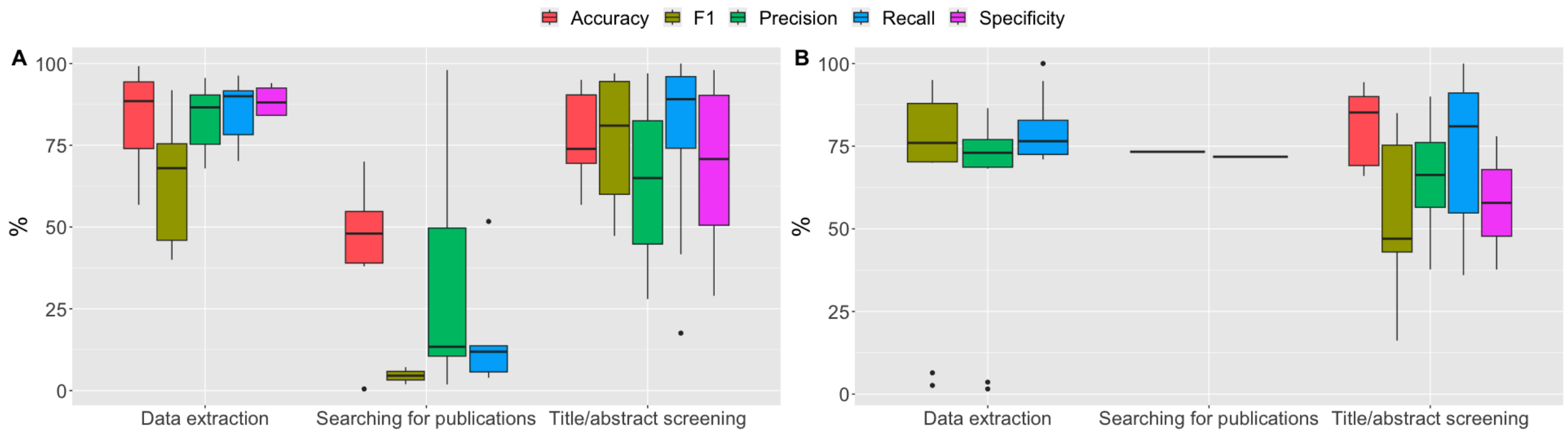

**Figure 5.** Performance metrics reported for the three most common automated stages **A:** for GPT-based models. **B:** for BERT-based models**.** *Source: ChatGPT-generated code using extracted data from this manuscript (Supplemental Table S7).*

Majority of reviewed publications were papers describing how LLM could be used to automate a certain phase of the review (n=146, 84.9%) (Supplemental Figure S6A). Only 26 (15.1%) papers were actual reviews conducted with some help from LLM tools. Majority of authors were positive about the usage of LLMs in reviews (n=120, 69.8%), with 43 citations (25.0%) containing mixed or cautious views on LLM usage (Supplemental Figure S6B). Only 9 (5.2%) study teams had negative experiences with LLM usage. Most studies had public funding reported (n=97, 56.4%) (Supplemental Figure S6C). When considering all the factors together, such as funding, journal impact factor, sample size, reported metrics, and others (see Methods), 72 citations (41.9%) appear to be of high quality, with 73 citations being medium quality (42.4%) (Supplemental Figure S6D).

Supplemental Table S7 presents the extraction table with all extracted categories across 172 citations.

## Discussion

Our LLM-assisted systematic review revealed a significant number of research projects related to review automation with LLM. Indeed, other researchers have noted promising results for LLMs in different areas, such as understanding human language and generating contextually appropriate responses [20-22].

Despite finding a significant number of projects using LLMs to automate some stages of the review process only few papers focused on the full cycle of review automation [23,24]. There might be perceived publication barriers, for example, journals recently started to ask about LLM-generated content, although we don't have information on whether this leads to changes in reviewing process. Growing number of LLM-generated papers will probably eventually change how review is conducted (reviewers might be assisted by LLMs or review paper format could be eventually replaced by online real-time information retrieval).

The strength of present review is in large-scale (over 3000 abstracts screened, and 172 full-text publications eligible for extraction) automation of different stages of review, including drafting the manuscript sections, and plot generation. Only few citations focused on automation of full cycle of review, while most focused only on specific areas like extraction or screening, including our own previous systematic review where GPT-3.5 was used with LDA-based topic modelling for validation of human findings [25]. In contrast, the LLM-based method that we applied in this work demonstrated its direct applicability, by facilitating the automation of the abstract and full-text screening, data extraction, as well as the knowledge synthesis stages, with the discussed constraints. Furthermore, our method is domain-agnostic, thus it can be integrated into large-scale review projects across different domains. The implications of such automation include reducing human workload and improving overall efficiency of systematic reviews. Furthermore, such tool in its more mature form will require less expertise from human reviewers, which could contribute to the democratization of systematic and scoping review process, with the potential to add features related to meta-analysis into the process.

GPT-based LLM were the most dominant type of LLM and the one that seems to show remarkable results on the data extraction, arguably the most complex and time-consuming stage of any review. It's usage for literature reviews is obvious, at this moment there are little restrictions on the type of information users can load into ChatGPT, and published papers are unlikely to contain any sensitive information, making ChatGPT with its high-performing model and developed API an obvious choice. At the same time smaller models like BERT, Llama or Mistral can be run and fine-tuned locally with much less cost, so we expect to see more automation projects with this LLM in the future.[26]

## Limitations

We used calibrated LLMs as reviewers in this project. Some extraction categories, such as performance metrics, had relatively lower accuracy, so the results of this extraction category should be taken with caution. Nevertheless, in this review LLMs achieved remarkable results in accuracy, making it possible to delegate time-consuming phases of review to LLMs. Studies generally recommend a single reviewer approach in some cases like rapid reviews[27]. However we believe that the LLM approach could substitute human reviewers, and human effort should be redirected to supervision of the review process.

A further limitation of this work is the simplified scoring system we introduced for research evaluation, which, using arbitrary weightings, may overlook key aspects like the novelty, robustness, and relevance of the studies. Future research should focus on improving LLM performance metrics, particularly precision and recall in lower-accuracy extraction categories. Additionally, integrating and evaluating different LLMs, possibly in combination with other AI models, should be explored to enhance performance. The short- and long-term impact of these integrations on review quality, along with ethical considerations, must also be assessed to maintain research credibility and trust.

## Conclusion

The use of LLMs in review automation is rapidly growing, with expected radical changes in scientific evidence synthesis. LLMs are likely to significantly reduce the time needed for reviews while producing similar or higher-quality data in greater quantities than manual reviews. Research shows it is becoming increasingly difficult to distinguish between LLM-generated and human-written text.[28] and the presence of LLM generated texts in scientific publications in growing exponentially [29]. To promote transparency and proper acknowledgment, researchers are encouraged to openly disclose their use of LLMs in academic papers, providing information on the prompts employed and the sections of text affected [30].

Despite early successes, few systematic reviews using LLMs were identified in our review. Although still in its early stages, AI-assisted reviews are already yielding impressive results, with growing interest as researchers develop semi-automated pipelines. However, generating trustworthy and useful AI-driven reviews still presents both technological and ethical challenges, particular for quantitative meta-analyses comparing treatment effects. However, the conduct of more simple systematic reviews, such as scoping reviews, appears to be well within the capabilities of current or near future AI methods.

**Author contributions**

LL, NH, and DS conceived and designed the review. DS developed the LLM screening automation add-on for Covidence. DS and VJ contributed to search strategy development. DS, AB and VJ performed the benchmarks for LLM and designed LLM prompts. DS, AB and VJ verified data extraction results. VJ and AB researched third-party components that were used to create the review. AB developed the script for journal impact factor assessment. DS analyzed the data and drafted the manuscript with the help of scite.ai and ChatGPT. NH and LL found the resources to conduct the review. All authors critically reviewed and revised the manuscript and approved the final version for submission.

**Funding**

This publication was supported, in part, by the National Center for Advancing Translational Sciences of the National Institutes of Health under Grant Number UL1 TR001450. Dr. Scherbakov was supported by grant T15 LM013977, Biomedical Informatics and Data Science for Health Equity Research (SC BIDS4Health). This publication was supported in part by a Smart-state Chair endowment. The content is solely the responsibility of the authors and does not necessarily represent the official views of the National Institutes of Health.

## Data availability

The data underlying this article are available in the article and supplementary materials.

## Conflicts of interest statement

The authors have no competing interests to declare.

## Supplementary Appendix

## Table of Contents

**Table S1. LLM Prompts used for screening and extraction.**

| Phase of the review | LLM prompt |
|---|---|
| Abstract screening | Summarise the text abstract of a full research paper (article), and given the below criteria list, say if the full paper is likely to be included, excluded, or unclear.<br>Definition of review. Review is a type of publication that synthesises knowledge from other publications.<br>Reviews include systematic, scoping reviews, meta-analysis, evidence synthesic, umbrella and rapid reviews, literature, narrative reviews, and other type of reviews.<br>A review typically has the following stages:<br>research question generation, creating a search strategy for a review, screening of literature, extraction of information, quality and bias assessment, evidence synthesis, writing a paper, generating code/plots for the review and generating tables.<br><br>Criteria list for exclusion/inclusion.<br>Include: Paper should be using some kind of large language models (LLM), like ChatGPT, GPT-3.5, GPT-4, Claude, BERT, BARD, Mistral, PaLM, Gemini, Copilot, Llama, Mixtral, and similar.<br>Include: Paper should be focused on automation of any stage of the review process listed above.<br>Exclude: If any of the Include criteria doesn't match.<br>Exclude: The paper is a review itself (types of review are listed above). However, if this review reports that it uses LLM for any review stage (stages of review are listed above), then include it.<br>Exclude: Paper is not related to automation of any parts of the review.<br>Exclude: Paper is a book chapter or compilation of conference papers (but single conference papers should be included).<br>Exclude: Paper mentions related technology like code generation with LLM but it is not related to creating a review (see definition of review above).<br>Exclude: Abstract and title are too brief and don't contain enough information to make the decision.<br>Follow this format:<br>1) First provide some explanations why each study should be included or excluded.<br>2) Then format your output as follows, strictly follow this format, use equal(=) sign, if study is excluded, write 'answer=excluded', if study is included output 'answer=included', or if it is unclear write 'answer=unclear'. |
| Full-text screening | Look at the research paper (article), and given the below criteria list, say if the full paper is to be included, excluded, or unclear.<br><br>Definition of review. Review is a type of publication that synthesises knowledge from other publications.<br>Reviews include systematic, scoping reviews, meta-analysis, evidence synthesic, umbrella and rapid reviews, literature, narrative reviews, and other type of reviews.<br>A review typically has the following stages:<br>research question generation, creating a search strategy for a review, review protocol creation, screening of literature, extraction of information, quality and bias assessment, evidence synthesis, writing a paper, generating code/plots for the review and generating tables.<br><br>Criteria list.<br>Paper should be focused on automation on any stage of the review with large language models (LLM).<br>If any of the following exclusion reason match, then exclude the article.<br>Exclude reason 1: Paper doesn't use some kind of large language models (LLM), like ChatGPT, GPT-3.5, GPT-4, Claude, BERT, BARD, Mistral, PaLM, Gemini, Copilot, Llama, Mixtral, and similar.<br>Exclude reason 2: Paper doesn't describe automation of any stage of the review process.<br>Exclude reason 3: Rather than covering automation of stages of the review process, paper is the review itself. However, if the paper is a review and uses some element of review automation to perform the review, then include it.<br>Exclude reason 4: Paper matches the focus (review automation with LLM), but it doesn't evaluate or report performance of any phase of the review process. Evaluation means verification by human experts. Often after evaluation performance metrics are reported which include, but not limited to: accuracy, F1, precision, recall, sensitivity, error rate, time saved, and others.<br>Exclude reason 5: Full text couldn't be retrieved.<br>Follow this format:<br>1) First provide some explanations why each study should be included or excluded.<br>2) Provide citation from text showing what NLP method was used and mental health problem explored.<br>3) Output the following:<br>include=yes/no/unclear<br>exclude_reason=reason_number (choose only one) |
| Extraction of data | Look at the research paper (article), and extract the following information. Follow the format given.<br>Definition of review. Review is a type of publication that synthesises knowledge from other publications.<br>Reviews include systematic, scoping reviews, meta-analysis, evidence synthetic, umbrella and rapid reviews, literature, narrative reviews, and other type of reviews.<br>A review typically has the following stages:<br>research question generation, creating a search strategy for a review, screening of literature, extraction of information, quality and bias assessment, evidence synthesis, writing a paper, generating code/plots for the review and generating tables.<br>Field 1) Extract country and US state (if it is in US) of this study. If this can not be determined from the text, look at the country and/or US state of first author's affiliation. Output as: Country name, or US/State Name<br>Field 2) What stages of the review process were automated in this study, choose all that apply: planning review, identifying research question, protocol creation, search strategy, searching for publications, abstract screening, full-text screening, extraction of data, bias or quality assessment, synthesis of knowledge, writing a paper, charts/plot generation, other(insert name). Output as comma separated string.<br>Field 3) What large language model (LLM) was used, for example: ChatGPT, GPT-3, GPT-3.5, GPT-4, BERT, BARD, Llama, Gemini, Mistral, Mixtral 8x7B, Claude, other [provide name and version].<br>Field 4) What were the performance metrics reported in the paper for best performing LLM model for each of the sage of review (Field 2)? Example output: Accuracy.Exctraction=0.84, F1.Abstract.Screening=0.23, Precision.FullText.Screening=0.4, Recall.FullTextScreening=0.3. Output as comma separated string.<br>Performance metrics include, but not limited to: accuracy, F1, precision, recall, sensitivity, error rate, specificity, positive predicted rate, kappa, inter-rater reliability and others.<br>Feild 5) For the metrics above, how were they calculated?<br>Field 6) How many papers/abstracts were used to estimate each of these metrics? Output as comma separated string. For example: Extraction=100 papers,Abstract.Screening=200 abstracts.<br>Field 7) What time saving was achieved if it was reported in this paper?<br>Field 8) What type of review is automated in this paper, select all that apply: systematic review, meta-analysis, scoping review, rapid review, narrative review, other ([insert name]). Output as comma separated string.<br>Field 9) What decision/conclusion authors make about the usage of LLMs in review automation? Is it positive, negative, or mixed? Provide citation from text to support your answer.<br>Field 10) What are academic ranks of authors in the paper? Output as comma separated string. For example: Postdoctoral scholar, Assistant Professor, Associate Professor.<br>Field 11) What are co-authors university affiliations? Output as comma separated string. For example: Medical University South Carolina, College of Charleston, Stanford University.<br>Field 12) Where does the funding come from? Select one of: public funding, private funding, or uknown sources. Output as comma separated string.<br>Format your output as an R data.frame:<br>data.frame(fld1='',fld2='',fld3='',...,fld12='') |

Source: Authors' own analysis

**Table S2. Benchmark of abstract screening phase (N=100 abstracts).**

| | Sensitivity | Specificity | Pos Pred Value | Neg Pred Value | Precision | Recall | F1 | Prevalence | Detection Rate | Detection Prevalence | Balanced Accuracy |
|---|---|---|---|---|---|---|---|---|---|---|---|
| Reviewer 1 vs Consensus | 0.77 | 0.98 | 0.97 | 0.82 | 0.97 | 0.77 | 0.86 | 0.48 | 0.37 | 0.38 | 0.88 |
| Reviewer 2 vs Consensus | 0.69 | 1 | 1 | 0.78 | 1 | 0.69 | 0.81 | 0.48 | 0.33 | 0.33 | 0.84 |
| Human consensus vs Consensus | 0.79 | 1 | 1 | 0.84 | 1 | 0.79 | 0.88 | 0.48 | 0.38 | 0.38 | 0.9 |
| LLM vs Consensus | 0.94 | 0.83 | 0.83 | 0.93 | 0.83 | 0.94 | 0.88 | 0.48 | 0.45 | 0.54 | 0.88 |

Source: Authors' own analysis

**Table S3. Benchmark of full-text screening phase (N=30 full-text PDFs).**

| | Sensitivity | Specificity | Pos Pred Value | Neg Pred Value | Precision | Recall | F1 | Prevalence | Detection Rate | Detection Prevalence | Balanced Accuracy |
|---|---|---|---|---|---|---|---|---|---|---|---|
| Reviewer 1 vs Consensus | 1 | 1 | 1 | 1 | 1 | 1 | 1 | 0.76 | 0.76 | 0.76 | 1 |
| LLM vs Consensus | 1 | 0.5 | 0.86 | 1 | 0.86 | 1 | 0.93 | 0.76 | 0.76 | 0.88 | 0.75 |

Source: Authors' own analysis

**Table S4. Benchmark of full-text extraction phase (N=15 full-text PDFs).**

| Category | Country | Review stage automated | LLM type used | Performance metrics of LLM | Sample size | Review type automated in the study | Authors opinion on LLM | Citation to support authors opinion | Type of funding used |
|---|---|---|---|---|---|---|---|---|---|
| Precision | 1 | 0.93 | 0.93 | 0.8 | 0.8 | 0.53 | 1 | 0.93 | 0.73 |
| Recall | 0.86 | 0.8 | 0.86 | 0.33 | 0.53 | 0.8 | 0.93 | 0.93 | 0.8 |

Source: Authors’ own analysis

**Time-saving and computational costs.**

Our review utilized approximately 500$ in OpenAI Azure costs for GPT-4o model.

We estimate that we saved time in screening 3241 abstracts (100 were manually screened for benchmark) by two reviewers with an average rate by a single reviewer of 40 abstracts per hour: 3241*2/40 = 162 hours. In addition, we saved time in screening 270 full-text publications (30 were manually screened for benchmark) by two reviewers with an average rate by a single reviewer of 10 full-texts per hour: 270*2/10 = 54 hours. We saved time in full-text extraction of 157 full-text publications (15 were manually extracted for benchmark) for 2 reviewers, but we had to do manual extraction of all papers for some categories where LLM precision/recall was low spending about 15 minutes per each publication, thus, assuming average rate of a single reviewer at 2 full-texts per hour we saved 157*2/2 – 157/4 = 118 hours. In addition, we saved time on drafting and code generation, with an estimated time saving of 50 hours.

Thus, we estimate total time saving of 334 person-hours.

**Figure S5. LLM model types used in the studies**

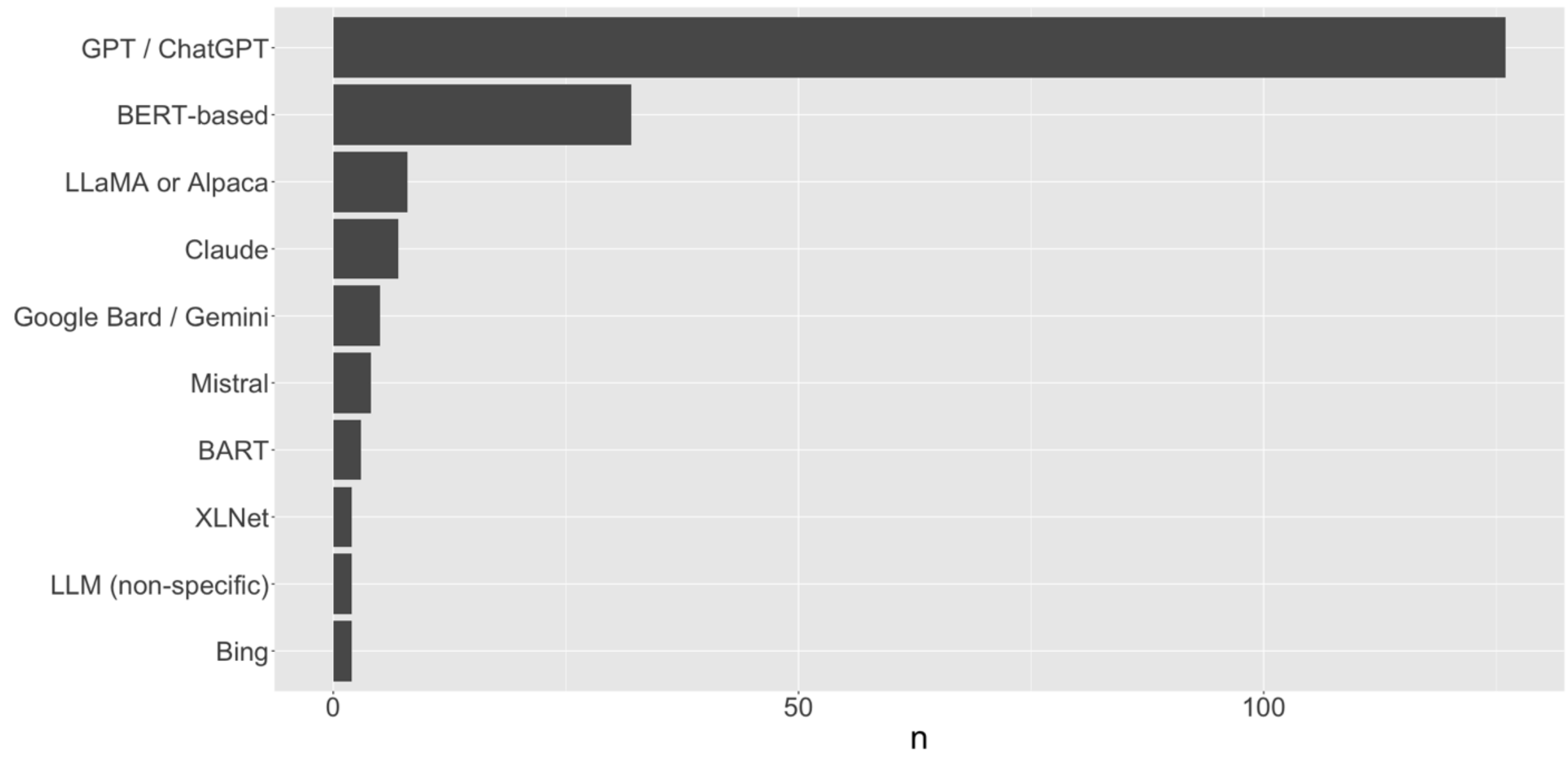

**Figure S5.** LLM types proposed for automation (models mentioned in 2 or more studies shown). Source: ChatGPT-generated code using extracted data

**Figure S6. Additional extracted categories.**

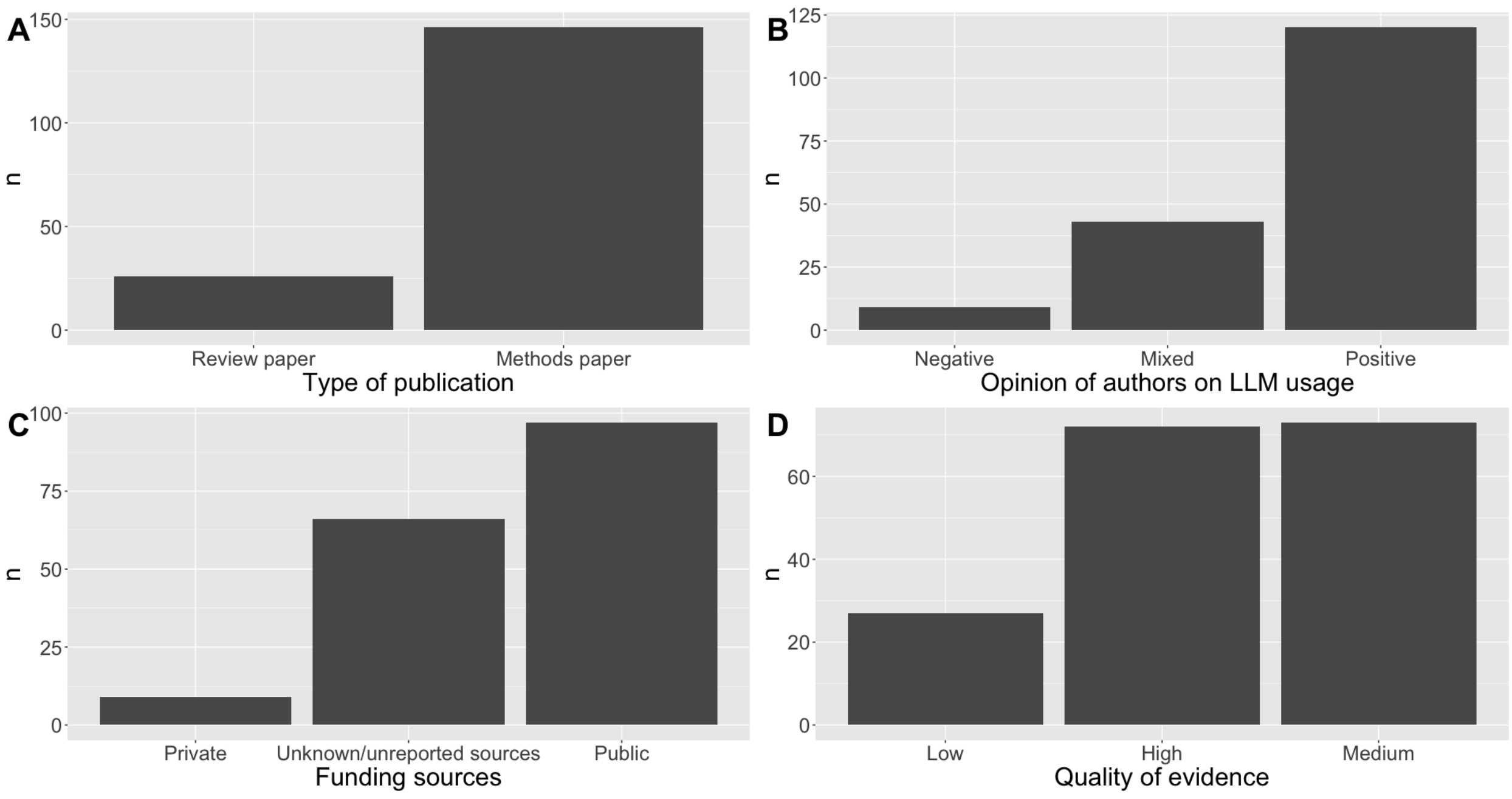

**Figure S6. A:** Type of citation (a review with LLM usage or a methods paper), **B:** Overall opinion of citation authors on LLM usage in review, **C:** Funding sources reported in the study, **D:** Overall quality of evidence**.** Source: ChatGPT-generated code using extracted data

**Table S7. Complete table of extracted categories.** † denotes categories that were verified by a human reviewer to ensure precision of extraction.

| Study | Title | Country/ US State | Review stage† | Review type† | LLM type† | Performance metrics† | Other metrics reported† | Details on performance metrics | Sample size† | Time savings reported | Review or methods study† | Fun-ding | Quality of evidence | Overall opinion† | Citation from study |
|---|---|---|---|---|---|---|---|---|---|---|---|---|---|---|---|
| **Guo, 2024 [1]** | Automated Paper Screening for Clinical Reviews Using Large Language Models: Data Analysis Study | Canada | Title and abstract screening | Systematic review, Scoping review | GPT / ChatGPT | GPT-4.Title and abstract screening.Accuracy=91.0; GPT-4.Title and abstract screening.F1=60.0 | Yes | Accuracy: computed by dividing papers selected by both GPT and human reviewers by the total number of papers. Macro F1-score: not specified in detail. Sensitivity: calculated for both included and excluded papers. Interrater reliability (kappa and PABAK): computed against the human-reviewed papers. | 24307 | Reduction in Screening Time with Gpt for the Noa Dataset Was Approximately 643 Minutes and Cost Approximately 25 | Methods paper | Unknown/unreported sources | High | Positive | "Large language models have the potential to streamline the clinical review process, save valuable time and effort for researchers, and contribute to the overall quality of clinical reviews." |
| **Haltaufderheide, 2024 [2]** | The Ethics of ChatGPT in Medicine and Healthcare: A Systematic Review on Large Language Models (LLMs) | Germany | Searching for publications | Rapid Review, Systematic review | GPT / ChatGPT | Not mentioned / Qualitative | No | Not extracted/Not applicable | 796 | Not extracted/Not applicable | Review paper | Public | High | Mixed | "Ethical examination of LLMs in healthcare is still nascent and struggles to keep pace with rapid technical advancements." |
| **Sun, 2024 [3]** | How good are large language models for automated data extraction from randomized trials? | China | Data extraction | Systematic review | ChatPDF, Claude | ChatPDF.Data extraction.Kappa =93.0; Claude.Data extraction.kappa=80.0 | Yes | Not extracted/Not applicable | 49 | Not extracted/Not applicable | Review paper | Public | High | Mixed | "Whilst promising, the percentage of correct responses is still unsatisfactory and therefore substantial improvements are needed for current AI tools to be adopted in research practice." |
| **Susnjak, 2023 [4]** | Prisma-dfllm: An extension of prisma for systematic literature reviews using domain-specific finetuned large language models | New Zealand | Searching for publications, Title and abstract screening, Full-text screening, Data extraction, Evidence synthesis/summarization | Systematic review, Systematic review | GPT / ChatGPT | Not mentioned / Qualitative | No | Not extracted/Not applicable | Not specified | Not extracted/Not applicable | Methods paper | Public | Medium | Positive | "The proposed extended PRISMA FLLM checklist of reporting guidelines provides a roadmap for researchers seeking to implement this approach." |
| **Susnjak, 2024 [5]** | Automating research synthesis with domain-specific large language model fine-tuning | New Zealand | Evidence synthesis/summarization, Data extraction | Systematic review | GPT / ChatGPT, Mistral | Not mentioned / Qualitative | Yes | not extracted | SYNTHESIS OF KNOWLEDGE=4962 | Not Extracted | Methods paper | Unknown/unreported sources | High | Positive | "AI technologies can effectively streamline SLRs, ensuring both efficiency and accuracy in information retrieval." |
| **Tang, 2023 [6]** | Evaluating large language models on medical evidence summarization | USA | Evidence synthesis/summarization | Meta-analysis, Systematic review | GPT / ChatGPT | Not mentioned / Qualitative | Yes | Performance metrics were calculated by comparing the generated summaries against reference summaries using ROUGE-L, METEOR, and BLEU scores, which measure overlap and precision of n-grams. | GPT-3.5.synthesis of knowledge=53 , ChatGPT.synthesis of knowledge=53 | Not Reported | Methods paper | Public | High | Negative | "Our study demonstrates that automatic metrics often do not strongly correlate with the quality of summaries ... LLMs could be susceptible to generating factually inconsistent summaries and making overly convincing or uncertain statements, leading to potential harm due to misinformation." |
| **Tran, 2024 [7]** | Sensitivity and Specificity of Using GPT-3.5 Turbo Models for Title and Abstract Screening in Systematic Reviews and Meta-analyses | France | Title and abstract screening | Rapid Review, Systematic review | GPT / ChatGPT | GPT-35 Turbo.Title and abstract screening.Recall=87.2; GPT-35 Turbo.Title and abstract screening.Specificity=52.2 | No | Comparing output of GPT-3.5 models under balanced and sensitive rules with original decisions from authors at title and abstract level, with sensitivities and specificities calculated using continuity corrected cell counts. | 22665 | Reducing the Number of Citations Before Manual Screening from 2 to 45 4 | Methods paper | Unknown/unreported sources | High | Mixed | "The GPT-3.5 Turbo model may be used as a second reviewer for title and abstract screening, at the cost of additional work to reconcile added false positives." |
| **Blasingame, 2024 [8]** | Evaluating a Large Language Model's Ability to Answer | USA/Tennessee | Evidence synthesis/summarization | Other/Non-specific | GPT / ChatGPT | Not mentioned / Qualitative | Yes | Not extracted/Not applicable | 216 | Not extracted/Not applicable | Methods paper | Public | High | Positive | "we envision this being the first of a series of investigations designed to |
| | Clinicians' Requests for Evidence Summaries | | | | | | | | | | | | | | further our understanding of how current and future versions of generative AI can be used and integrated into medical librarians workflow" |
| **Yan, 2023 [9]** | Leveraging Generative AI to Prioritize Drug Repurposing Candidates: Validating Identified Candidates for Alzheimer's Disease in Real-World Clinical Datasets | USA/Tennessee | Evidence synthesis/summarization | Meta-analysis | GPT / ChatGPT | Not mentioned / Qualitative | Yes | Calculated using Cox proportional hazards regression models comparing the risk of Alzheimers disease in individuals exposed to a drug repurposing candidate and propensity score-matched individuals never exposed to the drug | GPT-4.synthesis of knowledge=20 | Not Reported | Methods paper | Public | Medium | Positive | "Our findings suggest that ChatGPT can generate quality hypotheses for drug repurposing... With minimal costs, ChatGPT has the capacity and scalability to substantially accelerate the review process." |
| **Li, 2024 [10]** | Evaluating the Effectiveness of Large Language Models in Abstract Screening: A Comparative Analysis | USA/North Carolina | Title and abstract screening | Meta-analysis, Systematic review | GPT / ChatGPT, Google PaLM, Llama or Alpaca, Hybrid | ChatGPT4.Title and abstract screening.Accuracy=90.2; ChatGPT4.Title and abstract screening.Recall=89.1; ChatGPT4.Title and abstract screening.Specificity=90.7; ChatGPT35.Title and abstract screening.Accuracy=73.6; ChatGPT35.Title and abstract screening.Recall=74.1; ChatGPT4.Title and abstract screening.Specificity=78.0; Google PaLM.Title and abstract screening.Accuracy=78.6; Google PaLM.Title and abstract screening.Recall=49.9; Google PaLM.Title and abstract screening.Specificity=96.8; Meta Llama 2.Title and abstract screening.Accuracy=74.8; Meta Llama 2.Title and abstract screening.Recall=91.9; Meta Llama 2.Title and abstract screening.Specificity=65.7; Hybrid.Title and abstract screening.Accuracy=95.5; Hybrid.Title and abstract screening.Recall=53.9; Hybrid.Title and abstract screening.Specificity=98.4 | No | Sensitivity is defined as the number of true positives divided by the sum of true positives and false negatives, specificity as the number of true negatives divided by the sum of true negatives and false positives, and accuracy as sum of true positives and true negatives divided by the total number of abstracts. | 200 | Processing 200 Abstracts with Each Llm Took Approximately 10 20 Minutes using a Single Thread | Methods paper | Unknown/unreported sources | Medium | Mixed | "While LLM tools are not yet ready to completely replace human experts in abstract screening, they show great promise in revolutionizing the process." |
| **Wilkins, 2023 [11]** | Automated title and abstract screening for scoping reviews using the GPT-4 Large Language Model | Australia | Title and abstract screening | Scoping review | GPT / ChatGPT | GPT-4.Title and abstract screening.Accuracy=84.0; GPT-4.Title and abstract screening.Recall=71.0; GPT-4.Title and abstract screening.Specificity=89.0 | Yes | Accuracy was calculated as the proportion of correct decisions (both inclusions and exclusions) made by GPT-4 compared to the consensus human reviewer decision. Sensitivity was calculated as the proportion of true positives (correct inclusions) out of all actual positives (sources that should be included). Specificity was calculated | GPT-4.abstract screening=1147 | Not Reported | Methods paper | Public | High | Positive | "GPTscreenR demonstrates the potential for LLMs to support scholarly work and provides a user-friendly software framework that can be integrated into existing review pipelines." |
| | | | | | | | | as the proportion of true negatives (correct exclusions) out of all actual negatives (sources that should be excluded). | | | | | | | |
| **Oami, 2024 [12]** | Accuracy and reliability of data extraction for systematic reviews using large language models: A protocol for a prospective study | Japan | Data extraction | Systematic review | GPT / ChatGPT, Claude, Google Bard / Gemini | Not mentioned / Qualitative | Yes | Accuracy, F1, Precision, and Recall were calculated by comparing LLM-extracted data to a reference standard created by human reviewers. | Not extracted/Not applicable | Substantial Reduction in Time Compared to Conventional Methods Exact Time Savings not Reported | Methods paper | Unknown/unreported sources | High | Mixed | "This study aims to explore and evaluate the effectiveness of LLMs in systematic reviews, focusing on their potential to automate data extraction while ensuring high accuracy and minimal bias." |
| **Woelfle, 2024 [13]** | Benchmarking Human-AI Collaboration for Common Evidence Appraisal Tools | Switzerland, USA/California | Quality and bias assessment | Meta-analysis, Systematic review | Claude, GPT / ChatGPT, Mistral | Claude-3-Opus.Quality and bias assessment.Accuracy=70.0; Claude-2.Quality and bias assessment.Accuracy=70.0; GPT-4.Quality and bias assessment.Accuracy=69.0; GPT-35.Quality and bias assessment.Accuracy=63.0; Mixtral-8x22B.Quality and bias assessment.Accuracy=64.0; Claude-3-Opus.Quality and bias assessment.Accuracy=74.0; Claude-2.Quality and bias assessment.Accuracy=63.0; GPT-4.Quality and bias assessment.Accuracy=70.0; GPT-35.Quality and bias assessment.Accuracy=53.0; Mixtral-8x22B.Quality and bias assessment.Accuracy=59.0; Claude-3-Opus.Quality and bias assessment.Accuracy=45.0; Claude-2.Quality and bias assessment.Accuracy=44.0; GPT-4.Quality and bias assessment.Accuracy=38.0; GPT-35.Quality and bias assessment.Accuracy=55.0; Mixtral-8x22B.Quality and bias assessment.Accuracy=48.0 | Yes | Agreement with human consensus measured by accuracy (agreement fraction) and Cohens kappa. | Claude-3-Opus.bias or quality assessment=504, Claude-2.bias or quality assessment=504, GPT-4.bias or quality assessment=504, GPT-3.5.bias or quality assessment=504, Mixtral-8x22B.bias or quality assessment=504<br><br>Claude-3-Opus.bias or quality assessment=112, Claude-2.bias or quality assessment=112, GPT-4.bias or quality assessment=112, GPT-3.5.bias or quality assessment=112, Mixtral-8x22B.bias or quality assessment=112<br><br>Claude-3-Opus.bias or quality assessment=56, Claude-2.bias or quality assessment=56, GPT-4.bias or quality assessment=56, GPT-3.5.bias or quality | Not Reported | Methods paper | Public | High | Mixed | "Current LLMs alone appraised evidence worse than humans. Human-AI collaboration may reduce workload for the second human rater for the assessment of reporting (PRISMA) and methodological rigor (AMSTAR) but not for complex tasks such as PRECIS-2." |
| | | | | | | | | | assessment=56, Mixtral-8x22B.bias or quality assessment=56 | | | | | | |
| **Schmidt, 2024 [14]** | Exploring the use of a Large Language Model for data extraction in systematic reviews: a rapid feasibility study | United Kingdom | Data extraction | Systematic review | GPT / ChatGPT | GPT-4.Data extraction.Accuracy=80.0 | No | Each of the models responses was rated either complete, partial, or incorrect by two reviewers. If the models response contained all essential information or correctly did not provide a response when information was absent, it was rated complete. If some relevant information was present but missing other essential information, it was rated partial. Entirely incorrect or misleading responses were rated incorrect. | 100 | Not Reported | Methods paper | Public | Medium | Mixed | "Our results show that there might be value in using LLMs, for example as second or third reviewers. However, caution is advised when integrating models such as GPT-4 into tools." |
| **Yun, 2024 [15]** | Automatically Extracting Numerical Results from Randomized Controlled Trials with Large Language Models | USA/Massachusetts | Data extraction | Meta-analysis | GPT / ChatGPT, Llama or Alpaca, Mistral, Gemma, OLMo | GPT-4.Data extraction.F1=73.5; GPT-35.Data extraction.F1=68.0; Alpaca.Data extraction.F1=0.0; Mistral.Data extraction.F1=57.6; Gemma.Data extraction.F1=59.0; OLMo.Data extraction.F1=42.4; LLaMA.Data extraction.F1=12.4; BioMistral.Data extraction.F1=27.5 | Yes | Accuracy calculated as the proportion of exact matches; F1 calculated for binary and continuous outcomes; MSE calculated as the mean standardized error of the log odds ratio. | 172 | Not Reported | Methods paper | Public | Medium | Mixed | "The takeaway from this work is that modern LLMs offer a promising path toward fully automatic meta-analysis, but further improvements are needed before this will be reliable." |
| **Tsai, 2024 [16]** | Comparative Analysis of Automatic Literature Review Using Mistral Large Language Model and Human Reviewers | Taiwan | Searching for publications, Title and abstract screening, Full-text screening, Data extraction | Systematic review | Mistral | Not mentioned / Qualitative | Yes | Not extracted/Not applicable | 50 | Time Saving Was Reported as Mistral Llm Completing the Review Process in 17 Hours Compared to 100 Hours by Human Reviewers | Methods paper | Public | Medium | Mixed | "The findings indicate that while the Mistral LLM significantly surpasses human efforts in terms of efficiency and scalability, it occasionally lacks the analytical depth and attention to detail that characterize human reviews. Despite these limitations, the model demonstrates considerable potential in standardizing preliminary literature reviews." |
| **Robinson, 2023 [17]** | Bio-SIEVE: Exploring Instruction Tuning Large Language Models for Systematic Review Automation | United Kingdom | Title and abstract screening | Systematic review | GPT / ChatGPT, Llama or Alpaca, Guanaco | ChatGPT.Title and abstract screening.Accuracy=60.0; ChatGPT.Title and abstract screening.Precision=59.0; ChatGPT.Title and abstract screening.Recall=96.0; LLaMA.Title and abstract screening.Accuracy=74.0; LLaMA.Title and abstract screening.Precision=82.5; LLaMA.Title and abstract screening.Recall=71.5; Guanaco.Title and abstract screening.Accuracy=67.2; Guanaco.Title and abstract screening.Precision=72.5; Guanaco.Title and abstract screening.Recall=84.0 | No | Accuracy, Precision, and Recall were calculated based on the comparison of model predictions to the annotated labels in the test set. | ChatGPT.abstract screening=1001, LLaMA.abstract screening=1001, Guanaco.abstract screening=1001 | Not Reported | Methods paper | Public | High | Positive | "Bio-SIEVE lays the foundation for LLMs specialised for the SR process, paving the way for future developments for generative approaches to SR automation." |
| **Uittenhove, 2024 [18]** | Large Language Models in Psychology: Application in the Context of a Systematic Literature Review. | Switzerland | Data extraction | Systematic review | GPT / ChatGPT | GPT-4 turbo.Data extraction.Accuracy=95.0; GPT-4 turbo.Data extraction.Recall=96.2; GPT-4 turbo.Data extraction.Specificity=94.0; GPT-4 turbo.Data extraction.Accuracy=92.5; GPT-4 turbo.Data extraction.Recall=96.3; GPT-4 turbo.Data extraction.Specificity=84.2 | Yes | Cohens Kappa was calculated for inter-rater reliability. Sensitivity was calculated as TP / (TP + FN). Specificity was calculated as TN / (TN + FP). Accuracy was calculated as (TP + TN) / (TP + TN + FP + FN). The Area Under the ROC Curve (AUC) was also calculated. | extraction of data=39 articles | The Llm Completed Our Coding Tasks Significantly Faster than the Human Coders Taking Only a few Hours Compared to Several Days | Methods paper | Public | Medium | Positive | "Our results suggest that researchers and LLMs can work synergistically, improving efficiency, cost-effectiveness, and quality of the systematic literature review process." |
| **Wang, 2024 [19]** | MetaMate: Large Language Model to the Rescue of Automated Data Extraction for Educational Systematic Reviews and Meta-analyses | USA | Data extraction | Systematic review, Meta-analysis | GPT / ChatGPT | GPT-4 turbo.Data extraction.Precision=93.8; GPT-4 turbo.Data extraction.Recall=90.0; GPT-4 turbo.Data extraction.F1=91.8 | No | Precision, recall, and F1 score were calculated based on correctly extracted data (CED), missing data (MD), and incorrectly extracted data (IED). Precision = CED / (CED + IED), Recall = CED / (CED + MD), F1 Score = 2 * (Precision * Recall) / (Precision + Recall) | extraction of data=32 | Not Reported | Methods paper | Unknown/unreported sources | Medium | Positive | "These findings suggest that MetaMate could potentially replace or assist human coders in data extraction tasks, while maintaining or improving performance." |
| **Huotala, 2024 [20]** | The Promise and Challenges of Using LLMs to Accelerate the Screening Process of Systematic Reviews | Canada, Finland | Title and abstract screening | Systematic review | GPT / ChatGPT | GPT-35.Title and abstract screening.Precision=65.0; GPT-4.Title and abstract screening.Precision=50.0; GPT-35.Title and abstract screening.Recall=17.6; GPT-4.Title and abstract screening.Recall=41.7 | Yes | F1 and accuracy were calculated using standard formulas: F1 = 2 * (precision * recall) / (precision + recall), and accuracy = (true positives + true negatives) / total samples | abstract screening=20 | Not Reported | Methods paper | Public | Medium | Mixed | "Citation: Using LLMs for text simplification in the screening process does not significantly improve human performance. Using LLMs to automate title-abstract screening seems promising, but current LLMs are not significantly more accurate than human screeners." |
| **Yun, 2023 [21]** | Appraising the Potential Uses and Harms of LLMs for Medical Systematic Reviews | Australia, China, Greece, United Kingdom, USA | Drafting a publication | Systematic review | Galactica, BioMedLM, GPT / ChatGPT | Not mentioned / Qualitative | No | Qualitative analysis was conducted based on expert interviews to evaluate the outputs generated by the LLMs. | NA | Na | Methods paper | Public | Medium | Mixed | "Participants noted that LLMs are inadequate for producing medical systematic reviews directly given that they do not adhere to formal review methods and guidelines." |
| **Prasad, 2024 [22]** | Towards Development of Automated Knowledge Maps and Databases for Materials Engineering using Large Language Models | India | Data extraction | Systematic review | GPT / ChatGPT, Google Bard / Gemini | ChatGPT-35 turbo.Data extraction.F1=40.0; ChatGPT-35 turbo.Data extraction.F1=47.9; Google Gemini Pro.Data extraction.F1=50.0; Google Gemini Pro.Data extraction.F1=63.0 | Yes | F1 score was calculated using ROUGE metrics with the formula: 2 * (Precision * Recall) / (Precision + Recall). Exact Match and Relaxed Match were used to compute these values. | extraction of data=7 | Not Reported | Methods paper | Unknown/unreported sources | Medium | Positive | "Our method offers efficiency and comprehension, enabling researchers to extract insights more effectively." |
| **Serajeh, 2024 [23]** | LLMs in HCI Data Work: Bridging the Gap Between Information Retrieval and Responsible Research Practices | Iran, Italy | Data extraction | Other/Non-specific | GPT / ChatGPT, Llama or Alpaca | GPT35.Data extraction.Accuracy=58.0; LLama2.Data extraction.Accuracy=56.0; GPT35.Data extraction.meanabsoluteerror=7.0; Llama2.Data extraction.meanabsoluteerror=7.6 | No | Not extracted/Not applicable | 300 | Not extracted/Not applicable | Methods paper | Unknown/unreported sources | High | Positive | "This strategy not only ensured accuracy but also reduced surveillance risk." |
| **Wang, 2024 [24]** | Zero-shot Generative Large Language Models for Systematic Review Screening Automation | Australia, Germany | Title and abstract screening | Systematic review | GPT / ChatGPT, Llama or Alpaca | ChatGPT.Title and abstract screening.Recall=87.0; ChatGPT.Title and abstract screening.Recall=93.0; Llama.Title and abstract screening.Recall=89.0; Llama.Title and abstract screening.Recall=97.0; Alpaca.Title and abstract screening.Recall=91.0; Alpaca.Title and abstract screening.Recall=99.0 | Yes | Various performance metrics (B-AC, success rate, WSS) were computed across different datasets by comparing the predicted inclusion/exclusion against the ground truth labels. | abstract screening=600000 | Significant Screening Time Saved Compared to State of the Art Approaches Specific Time Savings not Quantified | Methods paper | Public | High | Positive | "Our comprehensive evaluation using five standard test collections shows that instruction fine-tuning plays an important role in screening, that calibration renders LLMs practical for achieving a targeted recall, and that combining both with an ensemble of zero-shot models saves significant screening |
| | | | | | | | | | | | | | | | time compared to state-of-the-art approaches." |
| **Cai, 2023 [25]** | Utilizing ChatGPT to select literature for meta-analysis shows workload reduction while maintaining a similar recall level as manual curation | The Netherlands | Title and abstract screening | Meta-analysis | GPT / ChatGPT | GPT35.screeningtitleandabstract.Precision=91.0; GPT4.screeningtitleandabstract.Precision=94.0; gpt4.screeningtitleandabstract.Recall=98.0; gpt35.screeningtitleandabstract.Recall=96.0; gpt35.screeningtitleandabstract.F1=94.0; gpt4.screeningtitleandabstract.F1=96.0 | No | Not extracted/Not applicable | 1000+ | Not extracted/Not applicable | Methods paper | Public | High | Positive | "We show here that its possible to have automatic selection of records for meta-analysis with ChatGPT by developing a pipeline named LARS" |
| **Tao, 2024 [26]** | GPT-4 Performance on Querying Scientific Publications: Reproducibility, Accuracy, and Impact of an Instruction Sheet | USA/California | Data extraction | Systematic review | GPT / ChatGPT | GPT-4.Data extraction.Accuracy=87.0; GPT-4.Data extraction.Recall=72.0; GPT-4.Data extraction.Precision=87.0 | No | Accuracy was defined as concordance between the correct answer and the GPT-4 response for Boolean and numerical questions. Recall was calculated as the proportion of true positives out of the sum of true positives and false negatives. Precision was calculated as the proportion of true positives out of the sum of true positives and false positives. F1 score was the harmonic mean of precision and recall: 2 x (recall * precision) / (recall + precision). | 3600 | The Overall Cost of using the Gpt 4 Api Was Significantly Reduced to Approximately Five Fold with the Release of Gpt 4 Turbo which is more Cost Effective but Exact Time Savings Were not Reported | Methods paper | Public | High | Positive | "GPT-4 possesses extensive knowledge about HIV drug resistance and it reproducibly answers Boolean, numerical, and list questions about HIV drug resistance papers. Its accuracy, recall, and precision of approximately 87%, 73%, and 87% without human feedback demonstrate its potential at performing this task." |
| **Tovar, 2023 [27]** | AI Literature Review Suite | USA/Tennessee | Searching for publications, Data extraction | Literature/Narrative review | GPT / ChatGPT, Llama or Alpaca | Not mentioned / Qualitative | No | Not extracted/Not applicable | Not extracted/Not applicable | Not extracted/Not applicable | Methods paper | Unknown/unreported sources | Low | Positive | "AI Literature Review Suite stands as a potent ally for researchers, enhancing efficiency and quality of scholarly endeavors while promoting accelerated innovation and progress" |
| **Tang, 2024 [28]** | Large Language Model in Medical Information Extraction from Titles and Abstracts with Prompt Engineering Strategies: A Comparative Study of GPT-3.5 and GPT-4 | China, Hong Kong SAR | Data extraction | Systematic review | GPT / ChatGPT | GPT-4.Data extraction.Accuracy=68.8; GPT-4.Data extraction.Accuracy=96.4; GPT-35.Data extraction.Accuracy=56.8; GPT-35.Data extraction.Accuracy=99.2 | No | Comparison of model outputs with ground truth using BERTScore, ROUGE-1, and a self-developed GPT-4 evaluator | 100 | 8 to 10 Hours of Human Labor Reduced to under 5 Minutes Gpt 3 5 or 40 Minutes Gpt 4 | Methods paper | Unknown/unreported sources | High | Positive | "Our result confirms the effectiveness of LLMs in extracting medical information, suggesting their potential as efficient tools for literature review." |
| **Kataoka, 2023 [29]** | Development of meta-prompts for Large Language Models to screen titles and abstracts for diagnostic test accuracy reviews | Japan | Title and abstract screening | Systematic review | GPT / ChatGPT | GPT35.Title and abstract screening.Recall=98.0; GPT4.Title and abstract screening.Recall=98.0; GPT35.Title and abstract screening.Specificity=29.0; GOT4.Title and abstract screening.Specificity=43.0 | No | Not extracted/Not applicable | Not extracted/Not applicable | Not extracted/Not applicable | Methods paper | Public | High | Positive | "Our study indicates that GPT-3.5 turbo can be effectively used to classify abstracts for DTA systematic reviews." |
| **Aronson, 2023 [30]** | Preparing to Integrate Generative Pretrained Transformer Series 4 models into Genetic Variant Assessment Workflows: Assessing Performance, Drift, and Nondeterminism Characteristics Relative to Classifying Functional Evidence in Literature | USA/Massachusetts | Full-text screening, Data extraction | Other/Non-specific | GPT / ChatGPT | GPT-4-Turbo.Data extraction.Recall=92.2; GPT-4-Turbo.Data extraction.Precision=95.6; GPT-4-Turbo.Data extraction.Specificity=92.0; GPT-4-Turbo.Data extraction.Recall=90.0; GPT-4-Turbo.Data extraction.Precision=74.0; GPT-4-Turbo.Data extraction.Recall=70.2; GPT-4-Turbo.Data | Yes | Standard deviations are calculated based upon the metrics produced in the 20 runs. | sample size = 72 | Not Reported | Methods paper | Public | High | Mixed | "Our data provides support for the contention that GPT-4 Turbo can be used to improve clinical workflows but does not support its use in fully automated variant assessment." |
| | | | | | | extraction.Precision=86.6; GPT-4-Turbo.Data extraction.Recall=88.0; GPT-4-Turbo.Data extraction.Precision=76.6 | | | | | | | | | |
| **Mahmoudi, 2024 [31]** | A Critical Assessment of Large Language Models for Systematic Reviews: Utilizing ChatGPT for Complex Data Extraction | USA/Massachusetts | Data extraction | Systematic review | GPT / ChatGPT | GPT-4.Data extraction.Accuracy=72.0 | No | Calculated as the percentage of correct responses compared to manual screening results. | 10 | Not Reported | Methods paper | Public | Medium | Mixed | "We underscore LLMs utility in systematic reviews for basic, explicit data extraction but reveal significant limitations in handling nuanced, subjective criteria, emphasizing the current necessity for human oversight." |
| **Guler, 2023 [32]** | Artificial Intelligence Research in Business and Management: A Literature Review Leveraging Machine Learning and Large Language Models | Australia | Publication classification | Literature/Narrative review, Systematic review | GPT / ChatGPT | Not mentioned / Qualitative | No | Not extracted/Not applicable | Not extracted/Not applicable | 6974 | Methods paper | Unknown/unreported sources | Low | Positive | "Researchers can leverage our study to align their work with the prevalent topics and academic disciplines, increasing the likelihood of publication success in CABS journals and addressing essential research gaps in specific domains" |
| **Urrutia, 2023 [33]** | Deep Natural Language Feature Learning for Interpretable Prediction | Chile | Title and abstract screening, Data extraction | Systematic review | GPT / ChatGPT | ChatGPT.Title and abstract screening.Precision=74.0; ChatGPT.Title and abstract screening.Precision=97.0; ChatGPT.Title and abstract screening.Recall=52.0; ChatGPT.Title and abstract screening.Recall=96.0; ChatGPT.Title and abstract screening.F1=60.0; ChatGPT.Title and abstract screening.F1=97.0; ChatGPT.Title and abstract screening.Accuracy=73.0; ChatGPT.Title and abstract screening.Accuracy=95.0 | No | Metrics were calculated by comparing the outputs of the LLMs (ChatGPT) and NLLFG (BERT-like model) against a manually annotated set of examples by an expert. Precision, recall, and F1-score were used for each task. | abstract screening=1983 | Not Reported | Methods paper | Public | High | Positive | "The DT models are simple and fully interpretable, and significantly outperforms a LLM like ChatGPT, while reaching performance metrics competitive with a deep learning (black-box) model." |
| **Agarwal, 2024 [34]** | LitLLM: A Toolkit for Scientific Literature Review | Canada | Searching for publications, Drafting a publication | Literature/Narrative review | GPT / ChatGPT | Not mentioned / Qualitative | No | Not extracted/Not applicable | Not extracted/Not applicable | Not extracted/Not applicable | Methods paper | Public | Medium | Positive | "While our system shows promise as a helpful research assistant, we believe that their usage should be disclosed to the readers, and authors should also observe caution in eliminating any possible hallucinations." |
| **Srivastava, 2023 [35]** | A day in the life of ChatGPT as an academic reviewer: Investigating the potential of large language model for scientific literature review | USA, USA/Washington | Quality and bias assessment | Literature/Narrative review, Systematic review | GPT / ChatGPT | Not mentioned / Qualitative | No | Not extracted/Not applicable | 11 | Not Reported | Methods paper | Private | Low | Positive | "Our experiments demonstrate that ChatGPT can review the papers ... and provide insights into their potential for acceptance or rejection." |
| **Ali, 2024 [36]** | Can machine learning help accelerate article screening for systematic reviews? Yes, when article separability in embedding space is high | Singapore | Title and abstract screening | Systematic review | GPT / ChatGPT | Not mentioned / Qualitative | Yes | WSS @ 95% quantifies the percentage work saved compared to random sampling at 95% recall. MCC is calculated from the confusion matrix considering recall, precision, false positives, and false omissions. | abstract screening=661 to 6578 | 1 2 to 75 6 Work Saved at 95 Recall Average of 46 Work Saved | Methods paper | Unknown/unreported sources | High | Positive | "There was good evidence that the separability of clusters of relevant versus irrelevant articles in high-dimensional embedding space can strongly predict whether ML screening can help." |
| **Liu, 2023 [37]** | CoQuest: Exploring Research Question Co-Creation with an LLM-based Agent | USA | Searching for publications | Other/Non-specific | GPT / ChatGPT | Not mentioned / Qualitative | No | Participants rated the generated research questions based on novelty, value, surprise, and relevance using a 5-point Likert scale. | IdentifyingResearchQuestion=504 RQs | Not Reported | Methods paper | Unknown/unreported sources | Medium | Mixed | "Although recent research has demonstrated the potential of using smaller language models to generate novel RQs [42], there remains a lack of empirical understanding about how humans evaluate AI-generated RQs.'" |
| **Bersenev, 2024 [38]** | Replicating a High-Impact Scientific Publication Using Systems of Large Language Models | Canada | Drafting a publication | Systematic review | GPT / ChatGPT | Not mentioned / Qualitative | No | Not extracted/Not applicable | Not extracted/Not applicable | Not Reported | Methods paper | Public | Medium | Mixed | "It is not entirely clear to us whether the conclusions our system draws are indeed based on the analysis it conducted, or whether they are simply from the knowledge base intrinsically contained within GPT" |
| **Spillias, 2023 [39]** | Human-AI Collaboration to Identify Literature for Evidence Synthesis | Australia | Searching for publications | Scoping review | GPT / ChatGPT | ChatGPT.Searching for publications.kappa=63.0 | Yes | Not extracted/Not applicable | Not extracted/Not applicable | 1098 | Methods paper | Public | High | Positive | "We show that AI can provide avenues for broadening effectiveness of a systematic reviews search strategy and may omit less than 1% of relevant articles in an automated screen based on predetermined screening criteria - a rate which is similar to human experts conducting the same screen." |
| **Syriani, 2023 [40]** | Assessing the Ability of ChatGPT to Screen Articles for Systematic Reviews | Canada | Title and abstract screening | Systematic review | GPT / ChatGPT | GPT-35 Turbo.Title and abstract screening.Recall=74.1; GPT-35 Turbo.Title and abstract screening.Precision=32.4; GPT-35 Turbo.Title and abstract screening.Specificity=66.6; GPT-35 Turbo.Title and abstract screening.F1=47.3; GPT-35 Turbo.Title and abstract screening.Accuracy=70.3 | No | The performance metrics were calculated using standard classifier evaluation techniques. Recall and precision were calculated based on the true positives, false positives, and false negatives obtained from the screening task. F2 score was used to balance recall and precision with a higher weight on recall. MCC was used to evaluate the overall quality of the binary classifications. | 5222 | No Time Saving Metrics Reported | Methods paper | Public | High | Positive | "Our results indicate that ChatGPT is a viable option to automate the SR processes, but requires careful considerations from developers when integrating ChatGPT into their SR tools." |
| **Li, 2024 [41]** | ChatCite: LLM Agent with Human Workflow Guidance for Comparative Literature Summary | China | Evidence synthesis/summarization | Literature/Narrative review | GPT / ChatGPT | GPT35.Evidence synthesis/summarization.G-score=3.4; GPT4.Evidence synthesis/summarization.G-score=3.5 | Yes | Not extracted/Not applicable | 50 | Not extracted/Not applicable | Review paper | Public | High | Positive | "Additionally, the literature summaries generated by ChatCite can be directly used for drafting literature reviews." |
| **Zhu, 2023 [42]** | Hierarchical Catalogue Generation for Literature Review: A Benchmark | China | Searching for publications | Systematic review | BART, GPT / ChatGPT | Not mentioned / Qualitative | Yes | Not extracted/Not applicable | 7637 | Not extracted/Not applicable | Methods paper | Public | High | Positive | "Our extensive analyses verify the high quality of our dataset and the effectiveness of our evaluation metrics. We further benchmark diverse experiments on state-of-the-art summarization models like BART and large language models like ChatGPT to evaluate their capabilities" |
| **Akinseloyin, 2023 [43]** | A Novel Question-Answering Framework for Automated Abstract Screening Using Large Language Models | United Kingdom | Title and abstract screening | Systematic review | GPT / ChatGPT | Not mentioned / Qualitative | Yes | Not extracted/Not applicable | 31 datasets | Not extracted/Not applicable | Methods paper | Public | Medium | Positive | "LLMs demonstrated proficiency in prioritising candidate studies for abstract screening using the proposed QA framework" |
| **Chern, 2023 [44]** | FacTool: Factuality Detection in Generative AI -- A Tool Augmented Framework for Multi-Task and Multi-Domain Scenarios | China | Quality and bias assessment | Literature/Narrative review, Systematic review | GPT / ChatGPT | GPT-4.Quality and bias assessment.Accuracy=99.0; GPT-4.Quality and bias assessment.Recall=90.0; GPT-4.Quality and bias assessment.Precision=100.0 ; GPT-4.Quality and bias assessment.F1=95.0 | No | ROUGE and BERTScore metrics were used for Claim Extraction; Claim-Level and Response-Level F1 scores were used for KBQA, Code Generation, Math Problems, and Scientific Literature Review | 100 | Not Reported | Methods paper | Public | High | Positive | "FACTOOL powered by GPT-4 significantly outperforms the self-check baselines in scientific literature review." |
| **Sami, 2024 [45]** | System for systematic literature review using multiple AI agents: Concept and an empirical evaluation | Finland | Searching for publications, Title and abstract screening, Full- | Systematic review | LLM (non-specific) | Not mentioned / Qualitative | Yes | Accuracy was calculated by comparing the model-generated search strings and results with manually curated ones. Precision | sample size = 10 | The Model Significantly Reduced the Time Required for Conducting SLRs | Methods paper | Public | Medium | Positive | "The researchers expressed strong satisfaction with the proposed model and |
| | | | text screening, Data extraction | | | | | and recall were measured by the relevance of retrieved documents during searching and screening stages. F1 score was computed for full-text screening to balance precision and recall. Data extraction accuracy was validated by comparing extracted data points with a gold standard set. | | by Approximately 60 | | | | | provided feedback for further improvement." |
| **Kılıç, 2023 [46]** | A Semi-Automated Solution Approach Recommender for a Given Use Case: a Case Study for AI/ML in Oncology via Scopus and OpenAI | Denmark | Searching for publications, Title and abstract screening, Data extraction, Other stages | Other/Non-specific | GPT / ChatGPT | GPT-35.Data extraction.Precision=67.9; GPT-35.Data extraction.Recall=90.0; GPT-35.Data extraction.F1=77.4 | No | Precision = TP/(TP + FP), Recall = TP/(TP + FN), F1-score = (2 * Precision * Recall) / (Precision + Recall) | 55 | Sarbold Llm Completes the Whole Task in a few Hours Compared to a Week for a Manual Review | Methods paper | Public | Medium | Positive | "SARBOLD-LLM demonstrates successful outcomes across various domains, showcasing its robustness and effectiveness." |
| **Wang, 2023 [47]** | Can ChatGPT Write a Good Boolean Query for Systematic Review Literature Search? | Australia | Searching for publications | Rapid Review, Systematic review | GPT / ChatGPT | ChatGPT.Searching for publications.Precision=1.9; ChatGPT.Searching for publications.Precision=11.7 ; ChatGPT.Searching for publications.Recall=3.9; ChatGPT.Searching for publications.Recall=51.7; ChatGPT.Searching for publications.F1=1.9; ChatGPT.Searching for publications.F1=7.2 | Yes | The metrics were calculated based on retrieved PubMed IDs evaluated for relevance using abstract-level relevant assessment. | 72 review topics from CLEF TAR 2017 and 2018 datasets, and 40 topics from the Seed Collection | Not Reported | Methods paper | Unknown/unreported sources | High | Positive | "Overall, our study demonstrates the potential of ChatGPT in generating effective Boolean queries for systematic review literature search." |
| **Wang, 2023 [48]** | Generating Natural Language Queries for More Effective Systematic Review Screening Prioritisation | Australia, Germany | Title and abstract screening | Systematic review | GPT / ChatGPT, Llama or Alpaca | Not mentioned / Qualitative | Yes | The performance metrics were calculated using standard information retrieval evaluation metrics like MAP, recall at various cutoffs, and WSS, based on the ranking of documents for systematic review topics. | AbstractScreening=80 topics | Not Reported | Methods paper | Public | High | Positive | "Our best approach is not only viable based on the information available at the time of screening, but also has similar effectiveness to the final title." |
| **Goldfarb, 2024 [49]** | Barriers and Suggested Solutions to Nursing Participation in Research: A Systematic Review with NLP Tools (Preprint) | Israel | Searching for publications | Systematic review | GPT / ChatGPT | GPT4.Searching for publications.Precision=98.0 | No | Not extracted/Not applicable | 26627 | Not extracted/Not applicable | Methods paper | Unknown/unreported sources | High | Positive | "Using natural language processing tools is a promising approach for conducting systematic reviews" |
| **Zhao, 2024 [50]** | A Literature Review of Literature Reviews in Pattern Analysis and Machine Intelligence | China | Searching for publications | Other/Non-specific, Other/Non-specific, Systematic review | GPT / ChatGPT | Not mentioned / Qualitative | Yes | Evaluated based on prompt effectiveness for topic identification in literature review papers using ChatGPT, with the effectiveness measured by Normalized Edit Distance (NED). | not extracted | Not Reported | Review paper | Public | High | Mixed | "The observed differences suggest that most AI-generated reviews still lag behind human-authored reviews in multiple aspects." |
| **Pitre, 2023 [51]** | ChatGPT for assessing risk of bias of randomized trials using the RoB 2.0 tool: A methods study | Canada | Quality and bias assessment | Systematic review | GPT / ChatGPT | Not mentioned / Qualitative | Yes | Not extracted/Not applicable | 34 | Not extracted/Not applicable | Methods paper | Unknown/unreported sources | Medium | Negative | Not extracted |
| **Mao, 2024 [52]** | A Reproducibility Study of Goldilocks: Just-Right Tuning of BERT for TAR | Australia | Title and abstract screening | Systematic review | BERT-based | Not mentioned / Qualitative | Yes | Calculated as the proportion of relevant documents retrieved among the top R retrieved documents, where R is the total number of relevant documents for a given category or topic. The review cost is computed as the cumulative product of cost structure coefficients | abstract screening=117557, abstract screening=149404, abstract screening=218484, abstract screening=30521, | The Run Time for Bert Was Reduced from 18 Hours to Between 2 75 Minutes and 42 Minutes per Topic Depending on the Dataset Size | Methods paper | Public | High | Positive | "Our findings suggest that the search for the Goldilocks epoch is a laborious way of improving the effectiveness of BERT-based classifier models in TAR. Instead, we suggest that considering the tasks characteristics and identifying an appropriate pre-trained BERT-like backbone may be a simpler and more effective |
| | | | | | | | | and the corresponding document numbers. | abstract screening=41 996, abstract screening=31 639 | | | | | | way to achieve better effectiveness in TAR tasks." |
| **Aydın, 2022 [53]** | OpenAI ChatGPT Generated Literature Review: Digital Twin in Healthcare | Turkey | Evidence synthesis/summarization | Literature/Narrative review | GPT / ChatGPT | Not mentioned / Qualitative | No | Not extracted/Not applicable | Not quanitified | Not extracted/Not applicable | Methods paper | Unknown/unreported sources | Low | Mixed | "The results are promising; however, the paraphrased parts had significant matches when checked with the Ithenticate tool." |
| **Jafari, 2024 [54]** | Streamlining the Selection Phase of Systematic Literature Reviews (SLRs) Using AI-Enabled GPT-4 Assistant API | Iran | Searching for publications | Systematic review | GPT / ChatGPT | Not mentioned / Qualitative | No | Not extracted/Not applicable | 1499 | Not extracted/Not applicable | Methods paper | Unknown/unreported sources | Medium | Positive | "While it offers improvements to current manual approaches, its full potential is yet to be realized and will likely be reached through ongoing refinement and expansion into subsequent stages of the SLR process" |
| **Srivastava, 2023 [55]** | A Rapid Scoping Review and Conceptual Analysis of the Educational Metaverse in the Global South: Socio-Technical Perspectives | India | Searching for publications | Rapid Review, Scoping review | GPT / ChatGPT | Not mentioned / Qualitative | No | No metrics reported. | No metrics reported. | No Time Savings Reported | Review paper | Unknown/unreported sources | Medium | Mixed | "The use of Large Language Models (LLM) like ChatGPT in academics is a recent phenomenon under development with several ethical concerns." |
| **Kim, 2024 [56]** | Systematic Review on Healthcare Systems Engineering utilizing ChatGPT | South Korea | Searching for publications, Evidence synthesis/summarization | Systematic review | GPT / ChatGPT | Not mentioned / Qualitative | No | Not extracted/Not applicable | 9809 | Not extracted/Not applicable | Methods paper | Public | Medium | Positive | "The utilization of ChatGPT in academic reviews offers advantages in terms of rapid information acquisition and accessibility" |
| **Wang, 2022 [57]** | Neural Rankers for Effective Screening Prioritisation in Medical Systematic Review Literature Search | Australia, Germany | Title and abstract screening | Systematic review | BERT-based | Not mentioned / Qualitative | Yes | The recall and precision were calculated using standard formulas: Recall = TP / (TP + FN), Precision = TP / (TP + FP) | 100 | Not Reported | Methods paper | Public | High | Positive | "Our results show that BERT-based rankers outperform the current state-of-the-art screening prioritisation methods." |
| **Lam, 2024 [58]** | Concept Induction: Analyzing Unstructured Text with High-Level Concepts Using LLooM | Canada, USA | Publication classification | Literature/Narrative review, Literature/Narrative review | GPT / ChatGPT | Not mentioned / Qualitative | Yes | Coverage was calculated as the ratio of examples accurately classified by the generated concepts to the total number of examples. | 400 | Not Explicitly Reported | Methods paper | Public | High | Positive | "LLooM improves upon the quality and coverage of topic models and helps expert analysts to uncover novel insights even on familiar datasets." |
| **Pedroso-Roussado, 2023 [59]** | Investigating the Limitations of Fashion Research Methods in Applying a Sustainable Design Practice: A Systematic Review | Portugal | Evidence synthesis/summarization, Drafting a publication | Systematic review | GPT / ChatGPT | Not mentioned / Qualitative | Yes | The accuracy of each stage was assessed based on subjective evaluation and feedback from the author, with no specific quantitative metrics provided. | sample size = 49 | Not Reported | Review paper | Unknown/unreported sources | Medium | Positive | "The authors acknowledge the potential of ChatGPT but also highlight the limitations of current methodologies. Citation: This study may be helpful for policy making since it uncovers the handicaps of performing relevant research in the realm of fashion design and fashion industry focusing a more sustainable practice, supplemented with a prototype guidance from ChatGPT to allow a fast and reliable discourse under the scope of the objective of this study." |
| **Li, 2024 [60]** | Explaining Relationships Among Research Papers | USA/Texas | Drafting a publication | Literature/Narrative review, Systematic review | GPT / ChatGPT | Not mentioned / Qualitative | Yes | The performance metrics were calculated based on human evaluation scores provided by domain experts. Each expert scored the generated related work sections on various parameters including fluency, coherence, relevance to the target paper, relevance to the cited papers, factuality, | writing a paper=27 | Not Reported | Methods paper | Public | Medium | Positive | "Our results suggest that using LLMs like GPT-4 can significantly streamline the review process, making it more efficient while maintaining high accuracy and quality." |
| | | | | | | | | usefulness, writing style, and overall quality. | | | | | | | |
| **Ambalavanan, 2020 [61]** | Cascade Neural Ensemble for Identifying Scientifically Sound Articles | USA/Arizona | Full-text screening | Meta-analysis, Systematic review | BERT-based | SciBERT.Full-text screening.Precision=64.2; SciBERT.Full-text screening.Precision=66.9; SciBERT.Full-text screening.Recall=89.1; SciBERT.Full-text screening.Recall=96.2; SciBERT.Full-text screening.F1=74.2; SciBERT.Full-text screening.F1=76.4 | No | The metrics were calculated using 10-fold cross-validation, with micro averaging across all folds to obtain precision, recall, and F measure. | sample size = 50590 | Not Explicitly Reported | Methods paper | Public | High | Positive | "The overall observation was that SciBERT based models offer superior performance in identifying scientifically sound articles compared to the early neural network models or feature-engineered models, even when the dataset is highly imbalanced (up to 1 to 32, positive to negative ratio)." |
| **Grokhowsky, 2023 [62]** | Reducing knowledge synthesis workload time using a text-mining algorithm for research location and subtopic extraction from geographically dependent research publications | USA/North Carolina | Data extraction, Publication classification | Rapid Review | BART | BART.Publication classification.Accuracy=78.0; BART.Publication classification.Precision=71.0; BART.Publication classification.F1=71.0; BART.Data extraction.Accuracy=85.0; BART.Data extraction.Precision=85.0; BART.Data extraction.F1=92.0 | No | True positives, false positives, true negatives, and false negatives were calculated for each variable. Precision, accuracy, and F1-Measure were then computed based on these values. | 1073 | The Process Took less than 5 Minutes to Complete Compared to Manual Review Time Reductions of 55 63 Translating to Months of Saved Time | Methods paper | Public | High | Positive | "Workload time reduction was achieved by this process (i.e., geoparsing, subtopic clustering, topic grouping, and linear regression) as the process took less than 5 minutes to complete." |
| **Likhareva, 2024 [63]** | Empowering Interdisciplinary Research with BERT-Based Models: An Approach Through SciBERT-CNN with Topic Modeling | USA/California | Publication classification | Systematic review | BERT-based | SciBERT.Publication classification.Precision=70.0; SciBERT.Publication classification.Recall=74.0; SciBERT.Publication classification.F1=70.0 | No | Precision, Recall, and F1 scores were calculated by comparing model predictions to the ground truth labels in the dataset. Metrics were derived from confusion matrices generated for each label and verified using classification reports and ROC curves. | 6361 | Not Reported | Methods paper | Public | High | Positive | "Our experiments show significant reductions in misclassification and improvements in accuracy and efficiency compared to a standard BERT model." |
| **Kats, 2023 [64]** | Relevance feedback strategies for recall-oriented neural information retrieval | The Netherlands | Title and abstract screening | Other/Non-specific | BERT-based | Not mentioned / Qualitative | Yes | Not extracted/Not applicable | SearchingForPublications=300 abstracts | The Method can Reduce Review Effort Between 17 85 and 59 04 Given a Fixed Recall Target of 80 | Methods paper | Unknown/unreported sources | High | Positive | "Our results show that this method can reduce review effort between 17.85% and 59.04%, compared to a baseline approach (of no feedback), given a fixed recall target." |
| **Marshalova, 2023 [65]** | Automatic Aspect Extraction from Scientific Texts | Russia | Data extraction | Other/Non-specific, Scoping review | BERT-based | BERT.Data extraction.F1=92.0 | No | Not extracted/Not applicable | 200 | Not extracted/Not applicable | Methods paper | Unknown/unreported sources | High | Positive | "We show that there are some differences in aspect representation in different domains, but even though our model was trained on a limited number of scientific domains, it is still able to generalize to new domains, as was proved by cross-domain experiments." |
| **Janes, 2022 [66]** | Open Tracing Tools: Overview and Critical Comparison | Austria, Finland, Italy | Data extraction | Systematic review | GPT / ChatGPT | Not mentioned / Qualitative | No | Not extracted/Not applicable | Not extracted/Not applicable | The Study Reported a Time Saving of Approximately 50 in the Literature Review Process Due to Automation | Review paper | Public | Medium | Positive | "The use of ChatGPT significantly enhanced the efficiency and accuracy of several stages of the review process, enabling a more streamlined and effective workflow." |
| **Yang, 2024 [67]** | Automating biomedical literature review for rapid drug discovery: Leveraging GPT-4 to expedite pandemic response. | USA | Title and abstract screening | Rapid Review, Systematic review | GPT / ChatGPT | GPT-4.Title and abstract screening.Accuracy=93.0; GPT-4.Title and abstract screening.F1=88.0; GPT-4.Title and abstract screening.Recall=83.0; GPT-4.Title and abstract | No | Evaluated using strati<U+FB01>ed <U+FB01>ve-fold cross-validation, measuring accuracy, sensitivity, F1 score, precision, and speci<U+FB01>city | FullText.Screening=250 papers for SARS-CoV-2, 189 papers for Nipah | Not Reported | Methods paper | Public | High | Positive | "These results highlight the utility of ChatGPT in drug discovery and development and reveal their potential to enable rapid drug target identification during a |

| | | | | | | | | | | | | | | |
|---|---|---|---|---|---|---|---|---|---|---|---|---|---|---|
| | | | | | | screening.Specificity=98.0; GPT-4.Title and abstract screening.F1=74.0; GPT-4.Title and abstract screening.Specificity=75.0; GPT-4.Title and abstract screening.Specificity=91.0 | | | | | | | | pandemic-level health emergency." |
| **Barsby, 2024 [68]** | Pilot study on large language models for risk-of-bias assessments in systematic reviews: A(I) new type of bias? | United Kingdom | Quality and bias assessment | Systematic review | GPT / ChatGPT | Not mentioned / Qualitative | Yes | Comparing ChatGPT3.5 and ChatGPT4 decisions with human (gold standard) assessments | bias or quality assessment=15 papers | Not Reported | Methods paper | Unknown/unreported sources | Medium | Mixed | "Overall, ChatGPT demonstrated moderate agreement, and minor disagreement with gold standard (human) assessment. While encouraging, this suboptimal performance precludes us from recommending ChatGPT be used in real-world RoB assessment." |
| **Chelli, 2024 [69]** | Hallucination Rates and Reference Accuracy of ChatGPT and Bard for Systematic Reviews: Comparative Analysis. | France | Searching for publications | Systematic review | GPT / ChatGPT, Google Bard / Gemini | GPT-35.Searching for publications.Precision=9.4; GPT-35.Searching for publications.Recall=11.9; GPT-4.Searching for publications.Precision=13.4 ; GPT-4.Searching for publications.Recall=13.7; Bard.Searching for publications.Precision=0.0; Bard.Searching for publications.Recall=0.0 | Yes | The metrics (precision, recall, F1-score) were calculated by comparing the references generated by the LLMs with the original systematic review references. Precision was calculated as the proportion of relevant papers retrieved out of all papers retrieved by the LLMs. Recall was the proportion of relevant papers retrieved out of all relevant papers in the original systematic reviews. F1-score is the harmonic mean of precision and recall. | searching for publications=471 | Not Reported | Methods paper | Unknown/unreported sources | High | Negative | "Given their current performance, it is not recommended for LLMs to be deployed as the primary or exclusive tool for conducting systematic reviews. Any references generated by such models warrant thorough validation by researchers." |
| **Lai, 2024 [70]** | Assessing the Risk of Bias in Randomized Clinical Trials With Large Language Models. | China | Quality and bias assessment | Systematic review | GPT / ChatGPT, Claude | ChatGPT.Quality and bias assessment.F1=90.0; Clause.Quality and bias assessment.F1=91.0 | No | Not extracted/Not applicable | Not quantified | Not extracted/Not applicable | Methods paper | Public | High | Positive | "In this survey study of applying LLMs for ROB assessment, LLM 1 and LLM 2 demonstrated substantial accuracy and consistency in evaluating RCTs, suggesting their potential as supportive tools in systematic review processes" |
| **Gwon, 2024 [71]** | The Use of Generative AI for Scientific Literature Searches for Systematic Reviews: ChatGPT and Microsoft Bing AI Performance Evaluation. | South Korea, USA/Michigan, USA/Ohio | Searching for publications | Systematic review | GPT / ChatGPT, Bing | ChatGPT.Searching for publications.Accuracy=0.5; Bing.Searching for publications.Accuracy=4.0 | Yes | Precision and recall were calculated based on the number of relevant studies identified by the AI compared to the benchmark set by human experts. | ChatGPT=1287;Bing=48 | Not Reported | Methods paper | Private | Medium | Negative | "The results suggest that the use of ChatGPT as a tool for real-time evidence generation is not yet accurate and feasible. Therefore, researchers should be cautious about using such AI." |
| **Gue, 2024 [72]** | Evaluating the OpenAI's GPT-3.5 Turbo's performance in extracting information from scientific articles on diabetic retinopathy. | Singapore | Data extraction | Systematic review | GPT / ChatGPT | Not mentioned / Qualitative | Yes | Not extracted/Not applicable | 20 | Not extracted/Not applicable | Methods paper | Private | Low | Positive | "t OpenAI\s GPT-3.5 Turbo may be adopted to extract simple information that is easily located in the text, leaving more complex information to be extracted by the researcher" |
| **Ruksakulpiwat, 2024 [73]** | Assessing the Efficacy of ChatGPT Versus Human Researchers in Identifying Relevant Studies on mHealth Interventions for Improving Medication Adherence in Patients With Ischemic Stroke When Conducting Systematic Reviews: Comparative Analysis. | Thailand, USA/Ohio | Searching for publications | Systematic review | GPT / ChatGPT | ChatGPT.Searching for publications.Precision=77.0 | No | Not extracted/Not applicable | 334 | Not extracted/Not applicable | Methods paper | Public | High | Mixed | "Ultimately, the choice between human researchers and ChatGPT depends on the specific requirements and objectives of each review, but the collaborative synergy of both approaches holds the potential to advance evidence-based research and decision- |
| | | | | | | | | | | | | | | | making in the health care field" |
| **Ghosh, 2024 [74]** | AlpaPICO: Extraction of PICO frames from clinical trial documents using LLMs. | India | Data extraction | Systematic review | Llama or Alpaca | Llama 2.Data extraction.Precision=81.0; Llama 2.Data extraction.Recall=75.0; Llama 2.Data extraction.F1=78.0; Llama 2.Data extraction.Accuracy=64.0; Llama 2.Data extraction.Precision=64.0; Llama 2.Data extraction.Recall=50.0; Llama 2.Data extraction.F1=56.0; Llama 2.Data extraction.Accuracy=39.0 | No | Qualitative assessment based on empirical results. | 53397 | Not Reported | Methods paper | Public | High | Positive | "Our empirical results show that our proposed ICL-based framework produces comparable results on all the version of EBM-NLP datasets and the proposed instruction tuned version of our framework produces state-of-the-art results on all the different EBM-NLP datasets." |
| **Lan, 2024 [75]** | Automatic categorization of self-acknowledged limitations in randomized controlled trial publications. | USA/Illinois | Evidence synthesis/summarization | Systematic review | BERT-based | BERT.Evidence synthesis/summarization.F1=82.1 | No | Not extracted/Not applicable | 1090 | Not extracted/Not applicable | Methods paper | Public | High | Positive | "Automatic extraction of limitations from RCT publications could benefit peer review and evidence synthesis, and support advanced methods to search and aggregate the evidence from the clinical trial literature" |
| **Issaiy, 2024 [76]** | Methodological insights into ChatGPT's screening performance in systematic reviews. | Iran | Title and abstract screening | Systematic review | GPT / ChatGPT | GPT-35 turbo.Title and abstract screening.Recall=95.0; GPT-35 turbo.Title and abstract screening.Specificity=65.0; GPT-35 turbo.Title and abstract screening.Precision=28.0 | Yes | Metrics were calculated using statistical analyses, including the Kappa coefficient for inter-rater agreement, ROC curve plotting, AUC calculation, and bootstrapping for p-values and confidence intervals. | abstract screening=1198 | Chat Gpt Completed the Screening Process Within an Hour While g Ps Took an Average of 7 10 Days | Methods paper | Unknown/unreported sources | High | Positive | "ChatGPT shows promise in automating the article screening phase of systematic reviews, achieving high sensitivity and workload savings." |
| **Choueka, 2024 [77]** | ChatGPT in Urogynecology Research: Novel or Not? | USA/New York | Searching for publications | Systematic review | GPT / ChatGPT | ChatGPT 35.Searching for publications.Accuracy=54.0 | Yes | Accuracy was calculated by dividing the number of novel ideas (no prior SRs published on the topic) by the total number of ideas suggested. For general research (GR) novelty accuracy rate, it was performed by dividing the number of novel ideas (no prior publications of any type on the topic) by the total number of ideas suggested. | Identifying research question=50 | No Time Saving Reported | Methods paper | Unknown/unreported sources | Medium | Mixed | "ChatGPT may be helpful for identifying novel research ideas in urogynecology, but its accuracy is limited...our results reveal that ChatGPTs suggestions are not consistently accurate and should be carefully audited by those using it." |
| **Raja, 2024 [78]** | Automated Category and Trend Analysis of Scientific Articles on Ophthalmology Using Large Language Models: Development and Usability Study. | USA/Tennessee | Publication classification | Systematic review | BART | BART.Publication classification.Accuracy=86.0; BART.Publication classification.F1=85.0 | No | Not extracted/Not applicable | 1000 | Not extracted/Not applicable | Methods paper | Public | Medium | Positive | "The proposed framework achieves notable improvements in both accuracy and efficiency" |
| **Khraisha, 2024 [79]** | Can large language models replace humans in systematic reviews? Evaluating GPT-4's efficacy in screening and extracting data from peer-reviewed and grey literature in multiple languages. | Ireland | Title and abstract screening, Full-text screening, Data extraction | Systematic review | GPT / ChatGPT | GPT-4.Title and abstract screening.Accuracy=67.0; GPT-4.Title and abstract screening.Recall=42.0; GPT-4.Title and abstract screening.Specificity=92.0; GPT-4.Full-text screening.Accuracy=54.0; GPT-4.Full-text screening.Recall=38.0; GPT-4.Full-text screening.Specificity=69.0; GPT-4.Data extraction.Accuracy=82.0; GPT-4.Data | No | Sensitivity (TP / (TP + FN)), Specificity (TN / (TN + FP)), Accuracy ((TP + TN) / (TP + TN + FP + FN)) | Title and abstract screening=300 titles/abstracts, Full-text screening=150 full texts, Data extraction=30 full texts | Not Reported | Methods paper | Public | High | Mixed | "Although our findings indicate that, currently, substantial caution should be exercised if LLMs are being used to conduct systematic reviews, they also offer preliminary evidence that, for certain review tasks delivered under specific conditions, LLMs can rival human performance." |
| | | | | | | extraction.Recall=75.0; GPT-4.Data extraction.Specificity=84.0 | | | | | | | | | |
| **Demir, 2024 [80]** | Enhancing systematic reviews in orthodontics: a comparative examination of GPT-3.5 and GPT-4 for generating PICO-based queries with tailored prompts and configurations. | Turkey | Searching for publications, Other stages | Meta-analysis, Systematic review | GPT / ChatGPT | Not mentioned / Qualitative | Yes | Accuracy was measured using a six-point Likert scale for both search strategy and searching for publications. | generation of PICO elements=41, generation of Boolean queries=77, generation of keywords=77 | Not Reported | Methods paper | Unknown/unreported sources | Medium | Positive | "Both ChatGPT 3.5 and 4 can be pivotal tools for generating PICO-driven queries in orthodontics when optimally configured." |
| **Noe-Steinmüller, 2024 [81]** | Defining suffering in pain. A systematic review on pain-related suffering using natural language processing. | Germany, Israel, USA/New Hampshire | Data extraction, Evidence synthesis/summarization | Systematic review | GPT / ChatGPT | Not mentioned / Qualitative | No | Definitions generated by LLMs were compared qualitatively to the manually synthesized definition. | Definition=60 papers, Full.Text=111 articles | No Time Saving Reported | Review paper | Unknown/unreported sources | Medium | Positive | "To validate the integrative definition of pain-related suffering obtained with manual qualitative methods (Table 2), we conducted 2 analyses using large language models (LLM) to generate definitions of pain-related suffering." |
| **Abd-Alrazaq, 2024 [82]** | Machine Learning-Based Approach for Identifying Research Gaps: COVID-19 as a Case Study. | Qatar | Code and plots generation | Other/Non-specific | BERT-based | Not mentioned / Qualitative | No | Not extracted/Not applicable | 1121433 | Not extracted/Not applicable | Review paper | Unknown/unreported sources | Medium | Positive | "The proposed machine learning\x96based approach has the potential to identify research gaps in scientific literature." |
| **Gartlehner, 2024 [83]** | Data extraction for evidence synthesis using a large language model: A proof-of-concept study. | Austria, USA/North Carolina | Data extraction | Systematic review | Claude | Claude 2.Data extraction.Accuracy=96.3; Claude 2.Data extraction.F1=98.0; Claude 2.Data extraction.Recall=96.2 | No | Accuracy was calculated as the proportion of correctly extracted data items: (TP + TN) / (TP + FP + TN + FN) | 160 | Not Reported | Methods paper | Private | Medium | Positive | "Leveraging LLMs has the potential to substantially enhance the efficiency and accuracy of data extraction for evidence syntheses." |
| **Yan, 2024 [84]** | Leveraging generative AI to prioritize drug repurposing candidates for Alzheimer's disease with real-world clinical validation. | USA/Tennessee | Searching for publications | Meta-analysis | GPT / ChatGPT | Not mentioned / Qualitative | No | Not extracted/Not applicable | EHR of 235000 participants | Not extracted/Not applicable | Methods paper | Public | Medium | Positive | "s. These findings suggest GAI technologies can assimilate scientific insights from an extensive Internet-based search space, helping to prioritize drug repurposing candidates and facilitate the treatment of diseases." |
| **Hasan, 2024 [85]** | Integrating large language models in systematic reviews: a framework and case study using ROBINS-I for risk of bias assessment. | USA/Minnesota | Quality and bias assessment | Systematic review | GPT / ChatGPT | Not mentioned / Qualitative | Yes | Not extracted/Not applicable | not quantified | Not extracted/Not applicable | Methods paper | Unknown/unreported sources | Medium | Negative | "Considering the agreement level with a human reviewer in the case study, pairing artificial intelligence with an independent human reviewer remains required" |
| **Giunti, 2024 [86]** | Cocreating an Automated mHealth Apps Systematic Review Process With Generative AI: Design Science Research Approach. | Ireland | Code and plots generation | Systematic review | GPT / ChatGPT | Not mentioned / Qualitative | No | qualitative assessment based on effectiveness and efficiency in replicating the initial steps of the background studies | Not extracted/Not applicable | The Overall Cocreation Process Exercise Had a Total Duration of 4 Hours and 39 Minutes | Methods paper | Public | Medium | Mixed | "Using the results from the ChatGPT-generated script to fully automate the process would likely require further work refining the script, either by using the steps of the background studies to base the script or by providing clearer starting prompts for the generative AI. However, leveraging this approach as a means to advance work when the software developing team was otherwise engaged was useful." |
| **Reason, 2024 [87]** | Artificial Intelligence to Automate Network Meta-Analyses: Four Case Studies to Evaluate the Potential Application of Large Language Models. | United Kingdom | Data extraction | Meta-analysis, Systematic review | GPT / ChatGPT | GPT4.Data extraction.Accuracy=99.0 | No | Not extracted/Not applicable | 20 runs | Not extracted/Not applicable | Methods paper | Private | Low | Positive | "This study provides a promising indication of the feasibility of using current generation LLMs to automate data extraction, code generation and NMA result interpretation, which could result in signifcant time savings and reduce human error" |
| **Maniaci, 2024 [88]** | Is generative pre-trained transformer artificial intelligence (Chat-GPT) a reliable tool for guidelines synthesis? A preliminary evaluation for biologic CRSwNP therapy. | Italy | Searching for publications | Systematic review | GPT / ChatGPT | Not mentioned / Qualitative | Yes | Not extracted/Not applicable | 12 | Not extracted/Not applicable | Methods paper | Unknow n/unrep orted sources | Medium | Positive | "The application of AI in decision-making protocols and the creation of therapeutic algorithms for biologic drug selection, could ofer fascinating future prospects in the management of CRSwNP." |
| **Jenko, 2024 [89]** | An evaluation of AI generated literature reviews in musculoskeletal radiology. | United Kingdom | Evidence synthesis/summar ization, Drafting a publication | Literature/Narrati ve review | GPT / ChatGPT | Not mentioned / Qualitative | Yes | Likert scale ratings by two fellowship-trained radiologists on accuracy, comprehensiveness, and relevance | sample size = 7 | Not Reported | Methods paper | Public | Medium | Mixed | "Summaries produced by AI in its current state require careful human validation." |
| **Chaker, 2024 [90]** | Easing the Burden on Caregivers- Applications of Artificial Intelligence for Physicians and Caregivers of Children with Cleft Lip and Palate. | USA | Evidence synthesis/summar ization | Literature/Narrati ve review | GPT / ChatGPT | ChatGPT 35.Evidence synthesis/summarization.Ac curacy=69.0 | No | Accuracy was determined by comparing ChatGPT-generated answers to professional answers from senior Pediatric Plastic Surgeons. | sample size = 13 | Not Reported | Methods paper | Unknow n/unrep orted sources | Medium | Positive | "AI can assist in multiple traditional perioperative strategies to reduce caregivers and patient anxiety." |
| **Lozano, 2024 [91]** | Clinfo.ai: An Open-Source Retrieval-Augmented Large Language Model System for Answering Medical Questions using Scientific Literature. | USA/Calif ornia | Searching for publications, Title and abstract screening, Full-text screening, Data extraction, Evidence synthesis/summar ization, Drafting a publication | Systematic review | GPT / ChatGPT | GPT-35.Searching for publications.Precision=22.4 ; GPT-35.Searching for publications.Recall=5.7 | Yes | Precision and recall were calculated as follows: Precision = \|RET(D, k) <U+2229> REL(D, q)\| / \|RET(D, k)\|, Recall = \|RET(D, k) <U+2229> REL(D, q)\| / \|REL(D, q)\| | 200 | Not Reported | Methods paper | Public | High | Positive | "Our system is not merely copying and pasting information from an SR review. Instead, it demonstrates a genuine ability to understand and represent the information effectively, resulting in enhanced performance compared to comparable tools." |
| **Sanii, 2024 [92]** | Utility of Artificial Intelligence in Orthopedic Surgery Literature Review: A Comparative Pilot Study. | USA/Mich igan | Searching for publications | Literature/Narrati ve review | GPT / ChatGPT, Perplexity.AI | Not mentioned / Qualitative | Yes | Success rates were determined by comparing the number of valid articles identified by the AI programs with the number of articles the control arm produced. | searching for publications= 132 articles | Mean Total Search Time for Chat Gpt Was 57 3 Seconds Compared to 644 15 Seconds for the Control Arm | Methods paper | Unknow n/unrep orted sources | Medium | Negative | "The current iteration of ChatGPT cannot perform a reliable literature review, and Perplexity.AI is only able to perform a limited review of the medical literature. Any utilization of these open AI programs should be done with caution and human quality assurance to promote responsible use and avoid the risk of using fabricated search results." |
| **Hossain, 2024 [93]** | Using ChatGPT and other forms of generative AI in systematic reviews: Challenges and opportunities. | USA/Texa s | Searching for publications, Title and abstract screening, Full-text screening, Data extraction | Systematic review | LLM (non-specific) | Not mentioned / Qualitative | No | no metrics | no metrics | Not Reported | Methods paper | Unknow n/unrep orted sources | Low | Mixed | "Generative AI may not be reliable enough to complete major tasks within a review... Such problems, amongst many others, can limit the scope of using generative AI in systematic reviews and similar evidence synthesis activities... It is necessary to recognize the potential of generative AI and set standards on ethical use of this technology in research projects, including systematic reviews." |
| **White, 2023 [94]** | Sample size in quantitative instrument-based studies published in Scopus up to 2022: An artificial intelligence aided systematic review. | Peru | Title and abstract screening, Data extraction | Systematic review | Claude | Not mentioned / Qualitative | No | Usability metric was calculated by comparing the sample size extraction results from Claude and Claude-Instant AI tools. Discrepancies between the tools were excluded, and the percentage of non-discrepant results that were usable was calculated. | no metrics | Not Reported | Review paper | Private | Low | Positive | "This is one of the first studies to use AI tools to assist in the analysis for a systematic review study." |
| **Schopow, 2023 [95]** | Applications of the Natural Language Processing Tool ChatGPT in Clinical Practice: | Germany | Searching for publications, Title and abstract screening, Full- | Systematic review | GPT / ChatGPT | ChatGPT 35.Title and abstract screening.Recall=100.0; ChatGPT 35.Title and | Yes | Sensitivity=TP/(TP+FN), Specificity=TN/(TN+FP), Precision=TP/(TP+FP), Accuracy=(TP+TN)/(TP+ | 155 | Not Reported | Review paper | Unknow n/unrep orted sources | High | Positive | "Our findings underscore the potential of NLP models, including ChatGPT, in |
| | Comparative Study and Augmented Systematic Review. | | text screening, Data extraction, Drafting a publication | | | abstract screening.Specificity=50.0; ChatGPT 35.Title and abstract screening.Precision=65.2; ChatGPT 35.Title and abstract screening.Accuracy=74.2; ChatGPT 35.Title and abstract screening.Recall=100.0; ChatGPT 35.Title and abstract screening.Specificity=41.2; ChatGPT 35.Title and abstract screening.Precision=39.6; ChatGPT 35.Title and abstract screening.Accuracy=56.8 | | TN+FP+FN), Chance Hit Rate=(Sensitivity*Prevalence)+(Specificity*(1<U+2212>Prevalence)) | | | | | | | performing systematic reviews and other clinical tasks." |
| **Aiumtrakul, 2023 [96]** | Navigating the Landscape of Personalized Medicine: The Relevance of ChatGPT, BingChat, and Bard AI in Nephrology Literature Searches. | USA | Searching for publications | Systematic review | GPT / ChatGPT, Bing, Google Bard / Gemini | ChatGPT.Searching for publications.Accuracy=38.0 ; Bing Chat.Searching for publications.Accuracy=30.0 ; Bard AI.Searching for publications.Accuracy=3.0 | Yes | Accuracy was calculated by dividing the number of accurate references by the total number of references provided by each AI chatbot during the search and extraction stages. | sample size = 240 | Not Reported | Methods paper | Unknown/unreported sources | Medium | Mixed | "The outcomes of this investigation draw attention to inconsistent citation accuracy across the different AI tools evaluated. Despite some promising results, the discrepancies identified call for a cautious and rigorous vetting of AI-sourced references in medicine." |
| **Roberts, 2023 [97]** | Comparative study of ChatGPT and human evaluators on the assessment of medical literature according to recognised reporting standards. | United Kingdom | Quality and bias assessment | Systematic review | GPT / ChatGPT | Not mentioned / Qualitative | Yes | Not extracted/Not applicable | 30 | Not extracted/Not applicable | Methods paper | Public | Medium | Positive | "LLMs like ChatGPT can help automate appraisal of medical literature, aiding in the identification of accurately reported research" |
| **Panayi, 2023 [98]** | Evaluation of a prototype machine learning tool to semi-automate data extraction for systematic literature reviews. | USA/Massachusetts | Data extraction | Systematic review | BERT-based | BERT.Data extraction.Precision=70.0; BERT.Data extraction.Recall=71.0; BERT.Data extraction.F1=70.0; BERT.Data extraction.Precision=74.0; BERT.Data extraction.Recall=72.0; BERT.Data extraction.F1=73.0 | No | Performance was measured using the F1 score, a metric that combines precision and recall. We defined successful matches as partial overlap of entities of the same type. | max=86;min =176 | Not Reported | Methods paper | Private | Medium | Positive | " With refinement, machine learning may assist with manual data extraction for SLRs." |
| **Whitton, 2023 [99]** | Automated tabulation of clinical trial results: A joint entity and relation extraction approach with transformer-based language representations. | United Kingdom | Data extraction | Systematic review | BERT-based | BioBERT.Data extraction.Precision=77.0; BioBERT.Data extraction.Recall=78.0; BioBERT.Data extraction.F1=78.0; sciBERT.Data extraction.Precision=77.0; sciBERT.Data extraction.Recall=81.0; sciBERT.Data extraction.F1=79.0; RoBERTa.Data extraction.Precision=72.0; RoBERTa.Data extraction.Recall=75.0; RoBERTa.Data extraction.F1=74.0 | No | Precision, recall, and F1 were calculated by comparing predicted labels to gold standard annotations. For NER, it was token-level exact match. For RE, it was exact match on entity pairs and their relations. Joint NER + RE combined both stages. Tabulation metrics used exact and relaxed matching criteria for tuples. | 595 | Not Reported | Methods paper | Public | High | Positive | "The final system is a proof of concept that the generation of evidence tables can be semi-automated, representing a step towards fully automating systematic reviews." |
| **Ng, 2023 [100]** | Semi-automating abstract screening with a natural language model pretrained on biomedical literature. | Singapore | Title and abstract screening | Systematic review | BERT-based | BERT.Title and abstract screening.Specificity=37.7; BERT.Title and abstract screening.Precision=37.7; BERT.Title and abstract | No | The metrics were calculated based on the comparison between the model predictions and the human reviewer decisions. | 14503 | 65 Reduction in the Number of Abstracts Screened by the Second Reviewer | Methods paper | Public | High | Positive | "incorporating it into the screening workflow, with the second reviewer screening only abstracts with conflicting decisions, translated into a |
| | | | | | | screening.F1=37.7; BERT.Title and abstract screening.Accuracy=70.2 | | Sensitivity was calculated as the number of abstracts included by both human reviewer and pBERT divided by the number of abstracts included by the human reviewer. Precision was calculated as the number of abstracts included by both human reviewer and pBERT divided by the number of abstracts included by pBERT. F1 score was calculated using the formula 2 * (precision * recall) / (precision + recall). Accuracy was calculated as the number of abstracts included by both human reviewer and pBERT divided by the total number of abstracts screened. Cohens Kappa was used to measure inter-rater agreement between the human reviewer and pBERT. | | | | | | | 65% reduction in the number of abstracts screened by the second reviewer." |
| **Teperikidis, 2024 [101]** | Prompting ChatGPT to perform an umbrella review. | Greece | Searching for publications, Title and abstract screening, Data extraction, Quality and bias assessment, Evidence synthesis/summarization, Drafting a publication | Umbrella Review | GPT / ChatGPT | Not mentioned / Qualitative | No | Comparison with human reviewers | Not extracted/Not applicable | Not Reported | Methods paper | Unknown/unreported sources | Low | Positive | "We believe that the introduction of such powerful language models in the field of evidence synthesis will revolutionise the way we process and assimilate medical knowledge." |
| **Khlaif, 2023 [102]** | The Potential and Concerns of Using AI in Scientific Research: ChatGPT Performance Evaluation. | Occupied Palestinian Territory | Drafting a publication | Systematic review, Literature/Narrative review | GPT / ChatGPT | Not mentioned / Qualitative | Yes | Plagiarism was calculated using plagiarism detection tools, checking for the percentage of text overlap with existing sources. | sample size = 4 articles, 50 abstracts | Not Reported | Methods paper | Public | Medium | Mixed | "ChatGPT has a strong potential to increase human productivity in research and can be used in academic writing. However, ChatGPT had a minor impact on developing the research framework and data analysis." |
| **Suppadungsuk, 2023 [103]** | Examining the Validity of ChatGPT in Identifying Relevant Nephrology Literature: Findings and Implications. | USA/Minnesota | Searching for publications | Literature/Narrative review | GPT / ChatGPT | ChatGPT-35.Searching for publications.Accuracy=42.0 | No | The metrics were calculated by comparing the references provided by ChatGPT against multiple reliable sources, such as PubMed, Google Scholar, and Web of Science, to verify their existence, relevance, and correctness. | 610 | Not Reported | Methods paper | Unknown/unreported sources | Medium | Negative | "Based on our findings, the use of ChatGPT as a sole resource for identifying references to literature reviews in nephrology is not recommended." |
| **LamHoai, 2023 [104]** | Comparing Meta-Analyses with ChatGPT in the Evaluation of the Effectiveness and Tolerance of Systemic Therapies in Moderate-to-Severe Plaque Psoriasis. | Belgium | Evidence synthesis/summarization | Meta-analysis, Meta-analysis | GPT / ChatGPT | Not mentioned / Qualitative | Yes | Not extracted/Not applicable | 28 | Not extracted/Not applicable | Methods paper | Unknown/unreported sources | Medium | Mixed | "ChatGPT can generate conclusions that are similar to MAs when the efficacy of fewer drugs is compared but is still unable to summarize information in a way that matches up to the results of MAs/NMAs when more than three molecules are compared." |
| **Rajjoub, 2024 [105]** | ChatGPT and its Role in the Decision-Making for the Diagnosis and Treatment of Lumbar Spinal Stenosis: A | USA/New York | Evidence synthesis/summarization | Literature/Narrative review | GPT / ChatGPT | Not mentioned / Qualitative | No | Not extracted/Not applicable | 40 | Not extracted/Not applicable | Review paper | Unknown/unreported sources | Low | Positive | "These results demonstrate the potential for implementing ChatGPT into the spine surgeon\s workplace as a |
| | Comparative Analysis and Narrative Review. | | | | | | | | | | | | | | means of supporting the decision-making process for LSS diagnosis and treatment” |
| **Mahuli, 2023 [106]** | Application ChatGPT in conducting systematic reviews and meta-analyses. | India | Data extraction, Quality and bias assessment | Meta-analysis, Systematic review | GPT / ChatGPT | Not mentioned / Qualitative | No | Not extracted/Not applicable | Not extracted/Not applicable | Not extracted/Not applicable | Methods paper | Public | Medium | Positive | "It shows promise in reducing workload and time, but careful implementation and validation are necessary." |
| **Dossantos, 2023 [107]** | Eyes on AI: ChatGPT's Transformative Potential Impact on Ophthalmology. | USA, Washington D.C. | Searching for publications, Evidence synthesis/summarization, Drafting a publication | Literature/Narrative review | GPT / ChatGPT | Not mentioned / Qualitative | No | Not specified | Not reported | Not Reported | Methods paper | Unknown/unreported sources | Low | Mixed | "ChatGPT can facilitate literature reviews, data analysis, manuscript development, and peer review, but issues of accuracy, bias, and ethics need careful consideration." |
| **Teperikidis, 2023 [108]** | Does the long-term administration of proton pump inhibitors increase the risk of adverse cardiovascular outcomes? A ChatGPT powered umbrella review. | Greece | Searching for publications, Title and abstract screening, Full-text screening, Data extraction, Drafting a publication | Umbrella Review | GPT / ChatGPT | Not mentioned / Qualitative | No | Not applicable | Not applicable | Not Reported | Review paper | Unknown/unreported sources | Medium | Positive | "Finally, ChatGPT was successfully prompted to execute most of the tasks involved in this review. We therefore feel that this tool will be of great assistance in the field of evidence synthesis in the near future." |
| **Anghelescu, 2023 [109]** | PRISMA Systematic Literature Review, including with Meta-Analysis vs. Chatbot/GPT (AI) regarding Current Scientific Data on the Main Effects of the Calf Blood Deproteinized Hemoderivative Medicine (Actovegin) in Ischemic Stroke. | Romania | Searching for publications | Meta-analysis, Systematic review | GPT / ChatGPT | Not mentioned / Qualitative | No | Not extracted/Not applicable | 37 | Not extracted/Not applicable | Review paper | Unknown/unreported sources | Low | Positive | "AI can provide valuable support in conducting PRISMA-type systematic literature reviews, including meta-analyse. There are limitations when using ChatGPT, particularly in distinguishing between truth and falsehood and determining the appropriateness of interpolation. Nevertheless, AI can provide valuable support in conducting PRISMA-type systematic literature reviews, including meta-analyses." |
| **Singh, 2023 [110]** | ChatGPT as a tool for conducting literature review for dry eye disease. | India | Searching for publications | Literature/Narrative review | GPT / ChatGPT | Not mentioned / Qualitative | No | Not extracted/Not applicable | 138 | Not extracted/Not applicable | Methods paper | Unknown/unreported sources | Low | Negative | "ChatGPT should not be used for literature reviews for dry eye disease as it could not retrieve appropriate articles reliably" |
| **Wu, 2023 [111]** | Addition of dexamethasone to prolong peripheral nerve blocks: a ChatGPT-created narrative review. | USA/New York | Drafting a publication | Literature/Narrative review | GPT / ChatGPT | Not mentioned / Qualitative | No | Not reported | Not extracted/Not applicable | Not Reported | Methods paper | Unknown/unreported sources | Medium | Negative | "At this time, we do not believe ChatGPT is able to replace human experts and is extremely limited in providing original, creative solutions/ideas and interpreting data for a subspecialty medical review article." |
| **Ruksakulpiwat, 2023 [112]** | Using ChatGPT in Medical Research: Current Status and Future Directions. | Thailand, USA/Ohio | Searching for publications | Literature/Narrative review, Systematic review | GPT / ChatGPT | Not mentioned / Qualitative | No | Not extracted/Not applicable | 114 | Not extracted/Not applicable | Review paper | Public | Medium | Mixed | "onalized medicine (16.67% each). Conclusion: ChatGPT has the potential to revolutionize medical research in various ways. However, its accuracy, originality, academic integrity, and ethical issues must be thoroughly discussed and improved before its widespread implementation in clinical research and medical practice" |
| **Liu, 2023 [113]** | How Good Is ChatGPT for Medication Evidence Synthesis? | USA/New York | Searching for publications, Title and abstract screening, Full-text screening, Data extraction, Evidence synthesis/summar | Other/Non-specific | GPT / ChatGPT | ChatGPT.Evidence synthesis/summarization.Precision=33.3; ChatGPT.Evidence synthesis/summarization.Recall=20.7; ChatGPT.Evidence synthesis/summarization.F1 | Yes | The metrics were calculated by comparing the summaries generated by ChatGPT and the proposed method against reference texts manually extracted from DrugBank using Rouge, BLEU, and | sample size = 10 | Not Reported | Review paper | Public | Medium | Mixed | "In light of these findings, it may be beneficial to use a combination of both summarization and neural language modeling methods to achieve a more comprehensive and accurate summary of the information." |
| | | | ization, Drafting a publication | | | =23.6; ChatGPT.Evidence synthesis/summarization.Precision=13.5; ChatGPT.Evidence synthesis/summarization.Recall=7.0; ChatGPT.Evidence synthesis/summarization.F1=8.2; ChatGPT.Evidence synthesis/summarization.Precision=31.7; ChatGPT.Evidence synthesis/summarization.Recall=19.5; ChatGPT.Evidence synthesis/summarization.F1=22.4 | | Levenshtein Distance scores. | | | | | | | |
| **Huang, 2023 [114]** | The role of ChatGPT in scientific communication: writing better scientific review articles. | Taiwan, USA/Alabama | Drafting a publication | Literature/Narrative review | GPT / ChatGPT | Not mentioned / Qualitative | No | Not Reported | Not Reported | Not Reported | Review paper | Public | High | Positive | "Overall, the use of AI tools like ChatGPT can significantly enhance both the efficiency and the quality of writing review articles for scientists." |
| **Qureshi, 2023 [115]** | Are ChatGPT and large language models "the answer" to bringing us closer to systematic review automation? | USA/Colorado | Searching for publications, Evidence synthesis/summarization, Other stages | Systematic review | GPT / ChatGPT | Not mentioned / Qualitative | No | The performance was evaluated based on the appropriateness of the output to the tasks. Errors were noted when output was unusable or required significant corrections. | Not extracted/Not applicable | Not Reported | Methods paper | Unknown/unreported sources | Low | Mixed | "ChatGPT and other LLMs hold promise in being integrated into systematic reviews, but they are not yet able to be used with confidence in any way." |
| **Temsah, 2023 [116]** | Overview of Early ChatGPT's Presence in Medical Literature: Insights From a Hybrid Literature Review by ChatGPT and Human Experts. | Saudi Arabia | Evidence synthesis/summarization | Literature/Narrative review | GPT / ChatGPT | Not mentioned / Qualitative | No | Not reported | sample size = 175 | Not Reported | Review paper | Public | Medium | Positive | "This hybrid approach allowed us to leverage the capabilities of ChatGPT in the review process while maintaining human oversight for quality and interpretation." |
| **Gupta, 2023 [117]** | Utilization of ChatGPT for Plastic Surgery Research: Friend or Foe? | USA/Michigan | Searching for publications | Systematic review | GPT / ChatGPT | Not mentioned / Qualitative | No | Qualitative assessment based on the novelty of systematic review ideas generated. | sample size = 80 | Not Reported | Methods paper | Unknown/unreported sources | Low | Positive | "Overall, we determined that ChatGPT was effective in forming novel systematic review ideas." |
| **Najafali, 2023 [118]** | Truth or Lies? The Pitfalls and Limitations of ChatGPT in Systematic Review Creation. | USA/Illinois | Searching for publications, Other stages, Drafting a publication | Systematic review | GPT / ChatGPT | Not mentioned / Qualitative | No | Novelty was assessed by checking systematic review ideas against popular search engines. | ResearchQuestion=240 systematic reviews | Not Reported | Methods paper | Public | Medium | Mixed | "ChatGPT in its current state is limited to generating ideas. There is a need for considerable improvements for it to be able to execute the entire systematic review process singlehandedly." |
| **Sallam, 2023 [119]** | ChatGPT Utility in Healthcare Education, Research, and Practice: Systematic Review on the Promising Perspectives and Valid Concerns. | Jordan | Searching for publications | Systematic review | GPT / ChatGPT | Not mentioned / Qualitative | No | Not extracted/Not applicable | 280 | Not extracted/Not applicable | Review paper | Public | Medium | Mixed | "hallucination, limited knowledge, incorrect citations, cybersecurity issues, and risk of infodemics" |
| **Gupta, 2023 [120]** | Expanding Cosmetic Plastic Surgery Research With ChatGPT. | USA/Missouri | Searching for publications | Systematic review | GPT / ChatGPT | ChatGPT.Searching for publications.Accuracy=55.0 | No | Accuracy was calculated by determining the percentage of systematic review ideas that were novel based on existing literature. | sample size = 240 | Not Reported | Methods paper | Unknown/unreported sources | Medium | Positive | "ChatGPT is an excellent tool that should be utilized by plastic surgeons." |
| **Martenot, 2022 [121]** | LiSA: an assisted literature search pipeline for detecting serious adverse drug events with deep learning. | France | Title and abstract screening, Full-text screening | Literature/Narrative review | BERT-based | BERT.Title and abstract screening.Precision=90.0; BERT.Title and abstract screening.Recall=81.0; BERT.Title and abstract screening.F1=85.0; BERT.Full-text screening.Precision=100.0; BERT.Full-text screening.Precision=85.0; BERT.Full-text | Yes | The performance metrics were calculated using a combination of annotated test sets and manual expert review. For abstract screening, precision and recall metrics were derived by comparing system predictions with human annotations. | abstract screening=448 abstracts, full-text screening=783 papers | Medical Reviewer Increases by a Factor of 2 5 the Number of Relevant Documents it can Collect and Evaluate Compared to a Simple Keyword Search | Methods paper | Public | High | Positive | "The use of LiSA therefore makes it possible to largely increase the volume of relevant papers found during a defined search time (by a factor 2.5), especially when serious ADRs mentions are rare in the literature." |
| | | | | | | screening.Recall=100.0; BERT.Full-text screening.Recall=67.0; BERT.Full-text screening.F1=93.0; BERT.Full-text screening.F1=80.0 | | | | | | | | | |
| **Wang, 2022 [122]** | PICO entity extraction for preclinical animal literature. | United Kingdom | Data extraction | Systematic review | BERT-based | BERT.Data extraction.F1=71.0; BERT.Data extraction.Recall=74.0; BERT.Data extraction.Precision=68.2 | No | F1 scores were calculated using the formula: F1 = 2 * (Precision * Recall) / (Precision + Recall). Precision and Recall were calculated as follows: Precision = (number of predicted correct entities) / (number of predicted entities), Recall = (number of predicted correct entities) / (number of true entities). | 400 | Not Reported | Methods paper | Public | High | Positive | "Our study indicates that of the approaches tested, BERT pre-trained on PubMed abstracts is the best for both PICO sentence classification and PICO entity recognition in the preclinical abstracts." |
| **Mutinda, 2022 [123]** | Automatic data extraction to support meta-analysis statistical analysis: a case study on breast cancer. | Japan | Data extraction | Meta-analysis | BERT-based | BERT.Data extraction.F1=70.0; BERT.Data extraction.F1=95.0 | Yes | The performance metrics such as Precision, Recall, and F1-score were calculated using the standard evaluation methods in the test set. | sample size = 1011 | Not Reported | Methods paper | Public | Medium | Positive | "Citation: 'The results show potential in automating the tasks and hope to increase interest in research on automating the entire integrated meta-analysis process.'" |
| **Buchlak, 2022 [124]** | Natural Language Processing Applications in the Clinical Neurosciences: A Machine Learning Augmented Systematic Review. | Australia, USA/Washington | Title and abstract screening | Systematic review | BERT-based, XLNet | BERT.Title and abstract screening.Accuracy=66.0; RoBERTa.Title and abstract screening.Accuracy=66.0; XLNet.Title and abstract screening.Accuracy=71.0 | No | Not extracted/Not applicable | 1131 | Not extracted/Not applicable | Review paper | Public | High | Positive | "As NLP technologies mature, the potential for them to generate clinical benefts for patients and providers grows. NLP and machine learning appear to be enhancing research and practice in the clinical neurosciences" |
| **Aum, 2021 [125]** | srBERT: automatic article classification model for systematic review using BERT. | South Korea | Title and abstract screening | Systematic review | BERT-based | BERT.Title and abstract screening.Accuracy=89.4; BERT.Title and abstract screening.Accuracy=94.3; BERT.Title and abstract screening.F1=66.1; BERT.Title and abstract screening.F1=78.5; BERT.Title and abstract screening.Precision=68.9; BERT.Title and abstract screening.Precision=83.3; BERT.Title and abstract screening.Recall=54.8; BERT.Title and abstract screening.Recall=91.1 | Yes | The performance metrics were calculated using standard evaluation techniques such as accuracy, precision, recall, F1 score, and area under the receiver operating characteristic curve (AUC). | sample size = 3268 and 409 | Not Reported | Methods paper | Unknown/unreported sources | High | Positive | "Our research shows the possibility of automatic article classification using machine-learning approaches to support SR tasks and its broad applicability." |
| **Wang, 2022 [126]** | Risk of bias assessment in preclinical literature using natural language processing. | United Kingdom | Quality and bias assessment | Systematic review | BERT-based | BERT.Quality and bias assessment.F1=34.4; BERT.Quality and bias assessment.F1=94.0; BERT.Quality and bias assessment.Recall=46.5; BERT.Quality and bias assessment.Recall=94.6; BERT.Quality and bias assessment.Precision=92.8; BERT.Quality and bias assessment.Precision=30.5; BERT.Quality and bias assessment.Specificity=75.1 ; BERT.Quality and bias assessment.Specificity=92.1 | No | Recall = True Positive / (True Positive + False Negative), Precision = True Positive / (True Positive + False Positive), F1 = (2 * Recall * Precision) / (Recall + Precision), Specificity = True Negative / (True Negative + False Positive) | Random Allocation=784 papers, Blinded Assessment Outcome=784 papers, Conflict of Interests=710 papers, Animal Welfare Regulations= 710 papers | Not Reported | Methods paper | Public | High | Positive | "Our models significantly outperform regular expressions for four risk of bias items." |
| **Buchlak, 2022 [127]** | Clinical outcomes associated with robotic and computer-navigated total knee arthroplasty: a machine learning-augmented systematic review. | Australia | Title and abstract screening | Systematic review | BERT-based, XLNet | BERT.Title and abstract screening.Accuracy=90.0; BERT.Title and abstract screening.Precision=55.0; BERT.Title and abstract screening.Recall=41.0; BERT.Title and abstract screening.F1=47.0; XLNet.Title and abstract screening.Accuracy=91.0; XLNet.Title and abstract screening.Precision=67.0; XLNet.Title and abstract screening.Recall=31.0; XLNet.Title and abstract screening.F1=42.0; RoBERTa.Title and abstract screening.Accuracy=90.0; RoBERTa.Title and abstract screening.Precision=58.0; RoBERTa.Title and abstract screening.Recall=36.0; RoBERTa.Title and abstract screening.F1=43.0 | No | Model classification performance was assessed using three-fold cross-validation, with accuracy, area under the receiver operating characteristic curve (AUC), precision, recall, F1 and Matthews correlation coefficient (MCC) metrics. | 456 | N a | Review paper | Unknown/unreported sources | High | Positive | "NLP shows promise for facilitating the systematic review process." |
| **Lu, 2021 [128]** | Revealing Opinions for COVID-19 Questions Using a Context Retriever, Opinion Aggregator, and Question-Answering Model: Model Development Study. | USA | Searching for publications, Evidence synthesis/summarization | Literature/Narrative review | BERT-based | Not mentioned / Qualitative | No | Not extracted/Not applicable | 47000 | Not extracted/Not applicable | Methods paper | Public | Medium | Positive | "The results demonstrate the usefulness of the proposed method in answering COVID-19\x96related questions with main opinions and capturing the trends of research about COVID-19 and other relevant strains of coronavirus in recent years" |
| **Qin, 2021 [129]** | Natural language processing was effective in assisting rapid title and abstract screening when updating systematic reviews. | China | Title and abstract screening | Systematic review | BERT-based | BERT.Title and abstract screening.Recall=96.0; BERT.Title and abstract screening.Specificity=78.0; BERT.Title and abstract screening.Accuracy=81.0 | No | Sensitivity and specificity were calculated by comparing the model prediction classification results of the text to the original classification labels of the text in the test set. | 947 | 64 1 Workload Reduction | Methods paper | Public | High | Positive | "NLP technology using the ensemble learning method may effectively assist in rapid literature screening when updating systematic reviews." |
| **Ambalavanan, 2020 [130]** | Using the contextual language model BERT for multi-criteria classification of scientific articles. | USA/Arizona | Title and abstract screening | Systematic review | BERT-based | SciBERT.Title and abstract screening.Precision=66.3; SciBERT.Title and abstract screening.Recall=87.2; SciBERT.Title and abstract screening.F1=75.3 | No | The metrics were calculated using 10-fold cross-validation on the Clinical Hedges dataset. | Abstract.Screening=49028 abstracts | Not Reported | Methods paper | Unknown/unreported sources | High | Positive | "Pre-trained neural contextual language models (e.g. SciBERT) performed well for screening scientific articles." |
| **Zhao, 2024 [131]** | Potential to transform words to watts with large language models in battery research | China | Searching for publications, Data extraction, Evidence synthesis/summarization, Drafting a publication, Code and plots generation | Literature/Narrative review | GPT / ChatGPT | Not mentioned / Qualitative | No | qualitative assessment based on demonstration and examples provided | 1000 | Not Explicitly Reported but Implied Significant Time Savings in Literature Review and Synthesis | Methods paper | Public | High | Positive | "The results underscore the versatility and proficiency of this research paradigm. While ChatGPT excels at addressing general inquiries, BatteryGPT goes a step further by drawing from domain knowledge to offer specific and expert-level responses, ultimately enriching the knowledge exploration and research experience in the realm of fast-charging technology." |
| **Edwards, 2024 [132]** | ADVISE: Accelerating the Creation of Evidence Syntheses for Global Development Using Natural Language Processing-Supported Human-Artificial Intelligence Collaboration | USA/Massachusetts | Title and abstract screening | Other/Non-specific | BERT-based | Not mentioned / Qualitative | Yes | Accuracy and F1 score were calculated at the default threshold of 0.5, using 85% of the papers in the training set to train the model and 15% as a validation set. These metrics were averaged over five runs. | abstract screening=68 539 | 68 5 Reduction in Human Screening Effort Compared to No Ai Assistance 16 8 Reduction Compared to Industry Standard and an Additional | Methods paper | Public | High | Positive | "These findings demonstrate how AI can accelerate the development of evidence synthesis products and promote timely evidence-based decision making in global development." |
| | | | | | | | | | | 30 Reduction with Active Learning | | | | | |
| **Ye, 2024 [133]** | A Hybrid Semi-Automated Workflow for Systematic and Literature Review Processes with Large Language Model Analysis | Australia | Title and abstract screening, Full-text screening, Data extraction | Literature/Narrative review, Systematic review | Google Bard / Gemini | Not mentioned / Qualitative | Yes | Accuracy was calculated by comparing the LLM output with human decisions for inclusion/exclusion and data extraction, and measuring the proportion of correct matches. | Abstract.Screening=390 abstracts, FullText.Screening=390 full-text articles, Extraction=390 articles | Not Explicitly Reported but Implied Through Reduction in Human Workload and Improved Time Efficiency | Methods paper | Public | High | Positive | "The hybrid workflow improved the accuracy of the case study by identifying 6/390 (1.53%) articles that were misclassified by the human-only process. It also matched the human-only decisions completely regarding the rest of the 384 articles. Given the rapid advances in LLM technology, these results will undoubtedly improve over time." |
| **Zamani, 2024 [134]** | Generative AI — The End of Systematic Reviews in PhD Projects? | Australia, New Zealand | Searching for publications, Title and abstract screening, Data extraction | Systematic review | GPT / ChatGPT | Not mentioned / Qualitative | No | Qualitative assessment based on human intervention and review. | Not extracted/Not applicable | The Overall Generative Ai Based Process Was at Least Four Times Faster than the Traditional Process to Produce the First Complete Drafts of an Sr Document | Methods paper | Unknown/unreported sources | Low | Positive | "Generative AI assistance delivered significant benefits in speed and quality (comprehensiveness) when conducting systematic reviews." |
| **Cambaz, 2024 [135]** | Use of AI-driven Code Generation Models in Teaching and Learning Programming: a Systematic Literature Review | The Netherlands | Searching for publications | Systematic review | GPT / ChatGPT, Codex | Not mentioned / Qualitative | No | Qualitative | 115 | Not extracted/Not applicable | Review paper | Public | Medium | Positive | "The use of LLM-based code generators in programming education presents a promising avenue with possibilities to improve student\s learning experience and alleviate the workload of teachers by providing assistance" |
| **Roy, 2024 [136]** | GEAR-Up: Generative AI and External Knowledge-based Retrieval Upgrading Scholarly Article Searches for Systematic Reviews | USA/South Carolina | Searching for publications | Systematic review | GPT / ChatGPT | Not mentioned / Qualitative | No | Not extracted/Not applicable | Not specified | Not extracted/Not applicable | Methods paper | Public | Medium | Positive | "Our system shows favorable reviews for reducing the librarian burden by providing highquality articles like a human librarian." |
| **Flaherty, 2024 [137]** | Beyond Plagiarism: ChatGPT as the Vanguard of Technological Revolution in Research and Citation | USA/New York | Searching for publications, Data extraction | Literature/Narrative review | GPT / ChatGPT | Not mentioned / Qualitative | No | N/A | Not extracted/Not applicable | N a | Methods paper | Unknown/unreported sources | Low | Positive | "ChatGPT empowers researchers with a tool that enhances collaboration, streamlines literature reviews, and assists in proper citation practices." |
| **Treviño-Juarez, 2024 [138]** | Assessing Risk of Bias Using ChatGPT-4 and Cochrane ROB2 Tool | Mexico | Quality and bias assessment | Systematic review | GPT / ChatGPT | Not mentioned / Qualitative | No | Not extracted/Not applicable | not reported | Not extracted/Not applicable | Methods paper | Unknown/unreported sources | Low | Positive | "With ChatGPT-4 and automation, evidence-based medicine is on the fast track to success." |
| **Wang, 2024 [139]** | When Young Scholars Cooperate with LLMs in Academic Tasks: The Influence of Individual Differences and Task Complexities | China | Evidence synthesis/summarization, Drafting a publication | Literature/Narrative review | GPT / ChatGPT | Not mentioned / Qualitative | No | N/A | N/A | N a | Methods paper | Unknown/unreported sources | Low | Mixed | Not extracted |
| **Anghelescu, 2023 [140]** | Chatgpt: "to be or not to be"… in academic research. the human mind's analytical rigor and capacity to discriminate between ai bots' truths and hallucinations | Romania | Searching for publications, Drafting a publication, Evidence synthesis/summarization | Literature/Narrative review, Systematic review | GPT / ChatGPT | Not mentioned / Qualitative | No | Qualitative assessment was done by comparing responses from March and September 2023, noting improvements and limitations in accuracy and relevance | Not extracted/Not applicable | Not Reported | Methods paper | Unknown/unreported sources | Low | Mixed | "ChatGPT might be a possible adjunct to academic writing and scientific research, considering any limitations that might jeopardize the study." |
| **Zimmermann, 2024 [141]** | Leveraging Large Language Models for Literature Review Tasks - A Case Study Using ChatGPT | Austria | Searching for publications | Systematic review | GPT / ChatGPT | ChatGPT.Searching for publications.Accuracy=70.0 | No | Not extracted/Not applicable | 585 | Not extracted/Not applicable | Methods paper | Public | Medium | Mixed | "We conclude that ChatGPT can support researchers to scan and evaluate literature by providing relatively accurate answers if provided with specific questions." |
| **Shah-Mohammadi, 2024 [142]** | Large Language Model-Based Architecture for Automatic Outcome Data Extraction to Support Meta-Analysis | USA/Utah | Data extraction | Meta-analysis | GPT / ChatGPT | GPT35.Data extraction.Accuracy=92.0 | No | Not extracted/Not applicable | 248 | Not extracted/Not applicable | Methods paper | Public | High | Positive | "Various prompts were crafted and iteratively refined to guide GPT in generating responses that accurately capture the core |
| | | | | | | | | | | | | | | | elements of clinical trials from the target papers.Positive, This approach could enhance efficiency, reduce human error, and potentially uncover patterns or relationships within vast datasets that might be challenging for manual methods." |
| **Whang, 2024 [143]** | ChatGPT for editors: enhancing efficiency and effectiveness | South Korea | Other stages | Literature/Narrative review | GPT / ChatGPT | Not mentioned / Qualitative | No | Not extracted/Not applicable | not specified | Not extracted/Not applicable | Methods paper | Private | Low | Positive | " This approach emphasizes that ChatGPT should be recognized not as a replacement for human judgment and expertise in editorial processes, but as a tool that plays a supportive and complementary role" |
| **Jain, 2024 [144]** | SciSpace Literature Review: Harnessing AI for Effortless Scientific Discovery | India | Searching for publications, Data extraction | Literature/Narrative review | LLM (non-specific) | Not mentioned / Qualitative | No | N/A | N/A | Not Reported | Methods paper | Private | Low | Positive | "The tool received an overall experience rating of 3.9/5, the quality of insights was rated at 3.8/5, and search results were rated at 4.1/5. This shows that the tool is effective in finding relevant scientific literature and also providing valuable insights." |
| **Atkinson, 2023 [145]** | ChatGPT and computational-based research: benefits, drawbacks, and machine learning applications | Australia | Code and plots generation, Evidence synthesis/summarization | Systematic review | GPT / ChatGPT | Not mentioned / Qualitative | No | Not extracted/Not applicable | Not extracted/Not applicable | Not extracted/Not applicable | Methods paper | Public | Medium | Positive | "Specifcally, it has illustrated how ChatGPT 3.5 can be used to review and refne, correct errors, and create new codes for research projects" |
| **Livberber, 2023 [146]** | Toward non-human-centered design: designing an academic article with ChatGPT | Turkey | Drafting a publication | Literature/Narrative review, Other/Non-specific | GPT / ChatGPT | Not mentioned / Qualitative | No | Not extracted/Not applicable | Not specified | Not extracted/Not applicable | Methods paper | Unknown/unreported sources | Low | Positive | "ChatGPT exhibits capabilities in aiding the design process, generating ideas aligned with the overall purpose and focus of the paper, producing consistent and contextually relevant responses to various natural language inputs, partially assisting in literature reviews, supporting paper design in terms of both content and format, and providing reasonable editing and proofreading for articles." |
| **Guo, 2023 [147]** | SciMine: An Efficient Systematic Prioritization Model Based on Richer Semantic Information | China | Searching for publications, Full-text screening | Systematic review | SPECTER | Not mentioned / Qualitative | Yes | Not applicable; only qualitative assessment reported | Not extracted/Not applicable | More than 10 Workload Saved Compared to the Current State of the Art | Methods paper | Unknown/unreported sources | High | Positive | "SciMine outperforms existing methods significantly, achieving the best-reported results in the literature." |
| **Scells, 2023 [148]** | Smooth Operators for Effective Systematic Review Queries | Germany | Data extraction, Other stages | Systematic review | BERT-based | BERT.Data extraction.Recall=83.4; BERT.Data extraction.Recall=71.2; BERT.Data extraction.Precision=3.6; BERT.Data extraction.Precision=1.6; BERT.Data extraction.F1=6.4; BERT.Data extraction.F1=2.6 | Yes | no calculations described | Not extracted/Not applicable | Not Reported | Methods paper | Public | High | Positive | "Using our smooth operator model, the effectiveness of existing systematic review literature search queries can be improved without changing the syntactic or semantic structure of queries." |
| **Alshami, 2023 [149]** | Harnessing the Power of ChatGPT for Automating Systematic Review Process: Methodology, Case Study, Limitations, and Future Directions | USA/Florida | Publication classification | Systematic review | GPT / ChatGPT | ChatGPT.Publication classification.Accuracy=88.0 | No | Not extracted/Not applicable | 496 journals | Not extracted/Not applicable | Methods paper | Public | High | Positive | "Notably, ChatGPT exhibits exceptional performance in filtering and categorizing relevant articles, leading to significant time and effort savings" |
| **Lamovšek, 2023 [150]** | Analysis of Research on Artificial Intelligence in Public Administration: Literature Review and Textual Analysis | Slovenia | Drafting a publication, Evidence synthesis/summarization | Literature/Narrative review, Literature/Narrative review | GPT / ChatGPT | Not mentioned / Qualitative | No | Qualitative assessment based on the interpretation of text by the authors in collaboration with GPT-4. | Sample size = 19 | No Time Saving Reported | Review paper | Public | Medium | Mixed | "The results of our study show that researchers equally report advantages and disadvantages of using AI in public administration." |
| **Semrl, 2023 [151]** | AI language models in human reproduction research: exploring ChatGPT's potential to assist academic writing | Austria | Searching for publications, Drafting a publication | Meta-analysis, Systematic review | GPT / ChatGPT | Not mentioned / Qualitative | No | Not extracted/Not applicable | 6 | Not extracted/Not applicable | Methods paper | Unknown/unreported sources | Low | Positive | "We advocate for open discussions within the reproductive medicine research community to explore the advantages and disadvantages of implementing this AI technology. Researchers and reviewers should be informed about AI language models, and we encourage authors to transparently disclose their use." |
| **Herbst, 2023 [152]** | Accelerating literature screening for systematic literature reviews with Large Language Models – development, application, and first evaluation of a solution | Germany | Searching for publications | Systematic review | GPT / ChatGPT | GPT-4.Searching for publications.falsepositives=2.0 | No | Not extracted/Not applicable | 2465 | Not extracted/Not applicable | Methods paper | Public | Medium | Positive | "Our initial results suggest a vast automation potential, despite some risks and limitations that have to be further navigated." |
| **Tang, 2023 [153]** | Guidance for Clinical Evaluation under the Medical Device Regulation through Automated Scoping Searches | Germany | Searching for publications | Scoping review | BERT-based | BERT.Searching for publications.Precision=73.3; BERT.Searching for publications.Recall=71.8 | Yes | The precision was calculated as the number of relevant documents retrieved by either method. Recall was evaluated using the CLEF 2018 eHealth TAR dataset, which provides a known set of relevant documents for each query. | sample size = 30 | Not Reported | Methods paper | Public | Medium | Positive | "Results indicate the potential of automated searches to provide device-specific relevant data from multiple databases while screening fewer documents than in manual literature searches." |
| **Miao, 2023 [154]** | Mining Topic Structure of AI Algorithmic Literature | China | Data extraction | Other/Non-specific | GPT / ChatGPT | Not mentioned / Qualitative | No | Qualitative assessment by domain experts | n/a | Not Reported | Methods paper | Public | Medium | Positive | Not extracted |
| **Antu, 2023 [155]** | Using LLM (Large Language Model) to Improve Efficiency in Literature Review for Undergraduate Research | USA/Oregon | Searching for publications, Evidence synthesis/summarization | Literature/Narrative review | GPT / ChatGPT | Not mentioned / Qualitative | No | Not extracted/Not applicable | not defined | Not extracted/Not applicable | Methods paper | Public | Low | Mixed | "we hypothesize that further development of human in the loop strategies may help mitigate these challenges and strengthen the potential of AI tools to support academic literature reviews." |
| **Twinomurinzi, 2023 [156]** | ChatGPT in Scholarly Discourse: Sentiments and an Inflection Point | South Africa | Publication classification | Scoping review | GPT / ChatGPT | Not mentioned / Qualitative | No | N/A | 67 | Not Reported | Review paper | Unknown/unreported sources | Low | Positive | "The key findings reveal a majority positive sentiment from scholars on ChatGPT mainly citing how academia should co-exist with the tool, and for researchers, organizations and society to use ChatGPT to stir greater creativity and productivity." |
| **Liang, 2023 [157]** | Sentiment analysis for software quality assessment | The Netherlands | Publication classification | Systematic review | BERT-based | BERT.Publication classification.Accuracy=80.0; BERT-BiLSTM.Publication classification.Accuracy=81.0; BERT-BiLSTM-Attention.Publication classification.Accuracy=82.0; RoBERTa.Publication classification.Accuracy=81.0 | No | Not extracted/Not applicable | Training=3107 reviews, Testing=1332 reviews | Not extracted/Not applicable | Methods paper | Public | Medium | Positive | " BERT-BiLSTM-Attention is selected as the sentiment analysis model due to its superior performance in both training and test datasets" |
| **Beheshti, 2023 [158]** | Transitioning drivers from linear to circular economic models: evidence of entrepreneurship in emerging nations | Czech Republic | Searching for publications, Data extraction | Systematic review | LangChain | Not mentioned / Qualitative | No | Qualitative assessment based on expert panel responses. | Not extracted/Not applicable | No Time Saving Reported | Review paper | Unknown/unreported sources | Medium | Mixed | Not extracted |
| **Platt, 2023 [159]** | Effectiveness of Generative Artificial Intelligence for Scientific Content Analysis | United Kingdom | Publication classification | Systematic review | text-bison | text-bison.Publication classification.Accuracy=90.0 | No | Not extracted/Not applicable | 41 | Not extracted/Not applicable | Methods paper | Unknown/unreported sources | Medium | Positive | "We conclude that some content analysis tasks with moderate accuracy requirements may be supported by current LLMs." |
| **Pattyn, 2023 [160]** | Preliminary Structured Literature Review Results Using ChatGPT: Towards a Pragmatic Framework for Product Managers at Software Startups | Belgium | Searching for publications, Data extraction | Systematic review | GPT / ChatGPT | Not mentioned / Qualitative | No | Consistency was calculated by ChatGPT 4.0 by validating and generating new task pairs, and the consistency rate was based on the number of change requests. | Sample size = 343 | No Specific Time Savings Reported | Review paper | Public | Medium | Positive | "ChatGPT has proven valuable for academics by streamlining research activities, saving time, and reducing uncertainty." |
| **Castillo-Segura, 2023 [161]** | Leveraging the Potential of Generative AI to Accelerate Systematic Literature Reviews: An Example in the Area of Educational Technology | Spain | Title and abstract screening | Systematic review | GPT / ChatGPT, Claude, LaMDA, Falcon-40b | Not mentioned / Qualitative | Yes | Metrics were calculated based on the confusion matrix with True Positive (TP), True Negative (TN), False Positive (FP), False Negative (FN), and Empty Result (ER). Additional performance metrics such as Precision, Sensitivity, Negative Prediction, Specificity, Accuracy, and F-measure were derived from these values. | sample size = 596 | Not Reported | Methods paper | Public | High | Mixed | "All in all, Forefront got the best overall performance (despite the high FP, which makes Spec and Prec low) followed by Bard and Claude." |
| **Ahmed, 2023 [162]** | Reimagining open data ecosystems: a practical approach using AI, CI, and Knowledge Graphs | Italy | Data extraction | Literature/Narrative review | BERT-based, GPT / ChatGPT | Not mentioned / Qualitative | Yes | Gestalt similarity score and Jaccard similarity score calculated between predicted and original keywords. | sample size = 1339 | Not Reported | Methods paper | Public | Medium | Positive | "In conclusion, the results of our research make a significant contribution towards enhancing the open data ecosystem, harnessing the potential of open data, and fostering innovation in various fields." |
| **Kartchner, 2023 [163]** | Zero-Shot Information Extraction for Clinical Meta-Analysis using Large Language Models | USA/Georgia | Data extraction | Meta-analysis | GPT / ChatGPT | ChatGPT.Data extraction.Accuracy=90.0; ChatGPT.Data extraction.F1=44.0 | No | Not extracted/Not applicable | 200 | Not extracted/Not applicable | Methods paper | Public | High | Positive | "The results of our research indicate that LLMs can contribute to more streamlined, transparent, and reproducible results in clinical research." |
| **Díaz, 2023 [164]** | Inquiry Frameworks for Research Question Scoping in DSR: A Realization for ChatGPT | Spain | Searching for publications | Scoping review | GPT / ChatGPT | Not mentioned / Qualitative | No | Not extracted/Not applicable | not reported | Not extracted/Not applicable | Methods paper | Public | Low | Positive | "This initial experience serves to identify three affordances for this kind of intervention: (1) an effective visualization to map out the research space to share with third parties (e.g., supervisors); (2) a search strategy to gradually narrow down the scope of the RQ to fit the resources available; and (3) a contextual state to keep a presence of the searching context throughout." |
| **Hasny, 2023 [165]** | BERT for Complex Systematic Review Screening to Support the Future of Medical Research | Germany | Title and abstract screening | Systematic review | BERT-based | BERT.Title and abstract screening.Recall=100.0; BERT.Title and abstract screening.Recall=69.3; BERT.Title and abstract screening.F1=43.9; BERT.Title and abstract screening.F1=16.2 | Yes | Recall, F1-Score, Screening Reduction, and AUC were calculated based on the models performance on the test dataset, comparing the predicted labels to the true labels. | sample size = 999 | No Time Saving Reported | Methods paper | Public | Medium | Positive | "Fine-tuning publicly available models, without the need for computationally expensive pretraining, scores recall values of at least 90% while reducing the human workload by at least 50%." |
| **Khadhraoui, 2022 [166]** | Survey of BERT-Base Models for Scientific Text Classification: COVID-19 Case Study | Tunisia | Publication classification | Systematic review | BERT-based | BERT.Publication classification.Accuracy=94.0; BERT.Publication classification.F1=86.0 | No | Not extracted/Not applicable | 4304 | Not extracted/Not applicable | Methods paper | Unknown/unreported sources | High | Positive | "The main approach is promising and presents an efficient increase based on the accuracy, precision, recall and F1 metrics." |
| **Shinde, 2022 [167]** | An Extractive-Abstractive Approach for Multi-document Summarization of Scientific Articles for Literature Review | USA | Data extraction, Evidence synthesis/summarization | Systematic review | BERT-based | BERT.Data extraction.F1=85.0 | No | Not extracted/Not applicable | 4500 | Not extracted/Not applicable | Methods paper | Unknown/unreported sources | High | Positive | "Although our results show that our hybrid approach can be used for generating fluent high-quality literature review summaries, there is still |
| | | | | | | | | | | | | | | | significant scope for improvement." |
| **Alchokr, 2022 [168]** | Supporting Systematic Literature Reviews Using Deep-Learning-Based Language Models | Germany | Searching for publications | Systematic review | BERT-based | Not mentioned / Qualitative | Yes | Not extracted/Not applicable | 174 | Not extracted/Not applicable | Methods paper | Unknown/unreported sources | Medium | Positive | "The findings indicate that using naturallanguage-based deep-learning architectures for semi-automating the selection of primary studies can accelerate the scanning and identification process." |
| **Yu, 2022 [169]** | Evaluating Pre-Trained Language Models on Multi-Document Summarization for Literature Reviews | USA/Georgia | Evidence synthesis/summarization | Systematic review | LongT5 | LongT5.Evidence synthesis/summarization.F1 =34.3 | No | Metrics such as Rouge-L, Rouge-1, Rouge-2, BERT score, EI (Evidence Inference), and F1 were calculated based on model outputs compared to benchmark datasets. These metrics assess the quality of generated summaries in terms of overlap with reference summaries and factual correctness. | 20000 | Not Reported | Methods paper | Unknown/unreported sources | High | Mixed | "We werent able to improve upon existing benchmarks for either the MS^2 or Cochrane datasets. We did show there is a need for stronger summarization metrics that can capture different linguistic dimensions such as factual correctness and readability." |
| **Yazi, 2021 [170]** | Towards Automated Detection of Contradictory Research Claims in Medical Literature Using Deep Learning Approach | Malaysia | Data extraction | Systematic review | BERT-based | BERT.Data extraction.Precision=80.7; BERT.Data extraction.Precision=86.5; BERT.Data extraction.Recall=94.7; BERT.Data extraction.Recall=100.0; BERT.Data extraction.F1=88.9; BERT.Data extraction.F1=92.2 | No | The performance of the models is measured using the precision, recall, and F1 score metrics, where the average results across 10-fold cross-validation were recorded. | sample size = 259 | Not Reported | Methods paper | Public | High | Positive | "the usage of deep learning approach can improve the detection of contradictory research claim in medical literature through the classification of claim assertion value compared to machine learning approach" |
| **Teslyuk, 2020 [171]** | The concept of system for automated scientific literature reviews generation | Russia | Evidence synthesis/summarization | Literature/Narrative review | BERT-based | Not mentioned / Qualitative | No | qualitative assessment based on the systems effectiveness in generating summaries | sample size = 643000 | Not Reported | Methods paper | Public | Medium | Mixed | "Currently the system is under our intensive development and testing."' |
| **Li, 2024 [172]** | RefAI: a GPT-powered retrieval-augmented generative tool for biomedical literature recommendation and summarization | USA | Searching for publications, Evidence synthesis/summarization | Systematic review | GPT / ChatGPT | Not mentioned / Qualitative | Yes | Metrics were calculated based on expert evaluations using a scale from 1 to 5, averaged for each tool across multiple subtopics. | couldn't be determined | No Time Savings Reported | Methods paper | Public | High | Positive | "RefAI demonstrated substantial improvements in literature recommendation and summarization over existing tools, addressing issues like fabricated papers, metadata inaccuracies, restricted recommendations, and poor reference integration." |

Source: GPT-4o analyzed content and authors' own analysis

**Appendix References**